\documentclass[12pt,preprint]{aastex}

\newcommand\UCHII{UCH\,{\sc ii}}
\newcommand\HII{H\,{\sc ii}}

\newcommand\HI{H\,{\sc i}}

\newcommand\kms{km~s$^{-1}$}

\newcommand\Msun{M$_{\odot}~$}

\newcommand\cmthree{cm$^{-3}~$}

\newcommand\etal{et al.~}
\newcommand\be{\begin{equation}}
\newcommand\ee{\end{equation}}
\newcommand\bea{\begin{eqnarray}}
\newcommand\eea{\end{eqnarray}}

\newcommand\ddeg{$^{o}$}

\catcode`\@=11
\newcommand{\gsim}{${\mathrel{\mathpalette\@versim>}}$}
\newcommand{\lsim}{${\mathrel{\mathpalette\@versim<}}$}
\newcommand{\@versim}[2]{\lower 2.9truept \vbox{\baselineskip 0pt \lineskip
    0.5truept \ialign{$\m@th#1\hfil##\hfil$\crcr#2\crcr\sim\crcr}}}
\catcode`\@=12

\slugcomment{}
\shorttitle{Helium Ionization in the \UCHII\ envelopes}
\shortauthors{}

\begin{document}

\title{Helium Ionization in the Diffuse Ionized Gas surrounding \UCHII\ regions}

\author{D. Anish Roshi}
\affil{National Radio Astronomy Observatory\altaffilmark{1},
520 Edgemont Road, Charlottesville, VA 22903, USA; aroshi@nrao.edu}

\author{E. Churchwell}
\affil{ Department of Astronomy, University of Wisconsin-Madison, 
475 N. Charter street, Madison, WI 53706, USA e-mail: churchwell@astro.wisc.edu}

\author{L. D. Anderson\altaffilmark{2}\altaffilmark{3}}
\affil{ Department of Physics and Astronomy, West Virginia University, Morgantown, WV 26506
USA e-mail: loren.anderson@mail.wvu.edu }

\altaffiltext{1}{The National Radio Astronomy Observatory is a facility of
the National Science Foundation operated under a cooperative
agreement by Associated Universities, Inc.}

\altaffiltext{2}{Faculty member at the Center for Gravitational Waves and Cosmology, West Virginia
University, Chestnut Ridge Research Building, Morgantown, WV 26505}
\altaffiltext{3}{Adjunct Astronomer at the Green Bank Observatory, PO Box 2, Green
Bank, WV 24944, USA}

\begin{abstract}
We present measurements of the singly ionized helium to hydrogen ratio ($n_{He^+}/n_{H^+}$) toward
diffuse gas surrounding three Ultra-Compact \HII\ (\UCHII ) regions: 
G10.15-0.34, G23.46-0.20 \& G29.96-0.02. We observe
radio recombination lines (RRLs) of hydrogen and helium near 5 GHz using the GBT
to measure the $n_{He^+}/n_{H^+}$ ratio. 
The measurements are motivated by
the low helium ionization observed in the warm ionized medium (WIM) 
and in the inner Galaxy diffuse ionized regions (DIR). 
Our data indicate that the helium is not uniformly ionized in the 
three observed sources. Helium lines are not detected 
toward a few observed positions in sources G10.15-0.34 \& G23.46-0.20 and 
the upper limits of the $n_{He^+}/n_{H^+}$ ratio obtained are 0.03 and 0.05 respectively.
The selected sources harbor stars of type O6 or hotter as indicated by
helium line detection toward the bright radio continuum emission from the sources
with mean $n_{He^+}/n_{H^+}$ value 0.06$\pm$0.02. 
Our data thus show that helium in diffuse gas located a few pc
away from the young massive stars embedded in the observed regions is not fully ionized.
We investigate
the origin of the non-uniform helium ionization 
and rule out the possibilities : (a)
that the helium is doubly ionized in the observed regions and (b) that
the low $n_{He^+}/n_{H^+}$ values are due to additional hydrogen ionizing 
radiation produced by accreting low-mass stars \citep{smith14}. 
We find that selective absorption of ionizing photons by dust can result
in low helium ionization but needs
further investigation to develop a self-consistent model for
dust in \HII\ regions.
\end{abstract}


\section{Introduction}
\label{sec:intro}

The existence of a diffuse ionized gas in the Galaxy is evident
from a variety of observations (\nocite{he63}Hoyle \& Ellis 1963,
see review by \nocite{hetal09}Haffner et al. 2009). This gas,
referred to as the Warm Ionized Medium (WIM), is now considered 
to be one of the major components of
the interstellar medium (ISM). The WIM has been primarily studied using
optical emission lines. These studies show that 
the local electron density of the WIM is in the range 0.01
to 0.1 \cmthree\ and emission measures are typically \lsim 10 pc cm$^{-6}$ 
\citep{hetal09}. In, or near the disk of the inner Galaxy, optical lines suffer
strong extinction and hence the distribution of the
ionized gas has been studied in the radio frequency regime.
In particular, low frequency (\lsim\ 2 GHz) radio recombination line (RRL) observations have 
detected diffuse ionized regions (DIR) with local density in the
range 1 to 10 \cmthree  and emission measure \lsim 1000 pc cm$^{-6}$
\citep{l76, a86, ra00, aetal15}. 
It is generally thought that both WIM and DIR 
are ionized by massive stars. To maintain ionization, the WIM and the DIR 
together require about 80\% of the ionizing radiation from all OB stars in
the Galaxy \citep{m78,mr10}. Thus the WIM and DIR form an 
energetically important component of the interstellar medium. 

The WIM in the Galaxy has been shown to have 
low $n_{He^+}/n_{H^+}$ number density ratios from optical line observations 
(\lsim 0.027; \citet{rt95}, see also \nocite{hetal09}Haffner \etal (2009)).
Here $n_{He^+}$ is the number density of singly ionized helium (He) and
$n_{H^+}$ is that of ionized hydrogen (H). RRL observations
toward DIR have provided an upper limit on $n_{He^+}/n_{H^+}$ of
$\sim$ 0.013 \citep{hetal96, retal12}. 
Ionization of both these components of the interstellar 
medium (ISM) is thought to be due to UV photons from O6-type or hotter stars  
that leak out of HII regions (\nocite{m78}Mezger 1978;
see also \nocite{andersonetal11}Anderson \etal 2011, \nocite{luisietal16}Luisi \etal 2016
for observational evidence of photon leakage from \HII\ regions).  
However, if stars hotter than $\sim$ O6 are the primary 
ionization sources of the DIR and the WIM, the ratio $n_{He^+}/n_{H^+}$ should be 
close to that of the actual He/H abundance ratio of $\sim$ 0.1 (see \nocite{d11}Draine 2011, 
Table 15.1 and section 15.5). This is because: (1) $\le$18\% of the ionizing 
photon flux from O6 or hotter stars, can fully ionize He in the ionized H region; 
(2) the ionization cross-sections 
for photons more energetic than the ionization potentials of H and He decrease with 
energy proportional to $(h\nu)^{-3}$ resulting in greater mean-free paths for the highest 
energy photons and expected ``hardening'' of the radiation field with distance from the source 
of ionization.  Thus, the low $n_{He^+}/n_{H^+}$ ratios in the WIM and DIR  
are not understood. 

There are at least two important issues related to the low 
helium ionization in the diffuse regions. Firstly, the observed $n_{He^+}/n_{H^+}$ ratio
toward several \HII\ regions harboring O6 or hotter stars is lower than the
cosmic abundance of helium \citep{setal83}. The mean value of $n_{He^+}/n_{H^+}$ ratio 
obtained toward such \HII\ regions is 0.08, indicating the presence of 20 \% neutral 
helium. The possible effects that can produce lower helium ionization
include (a) selective absorption by dust (\citet{msc74}; see Section~\ref{sec:dis}) 
and (b) line-blanketing in the atmosphere of the O stars 
(see for example \nocite{msh05}Martins, Schaerer \& Hiller 2005). 
Secondly and possibly even more problematic is how the ionizing UV photons escape through the 
HII ionization fronts that surround hot O stars and the surrounding DIRs of 
compact HII regions.  In particular, are the $n_{He^+}/n_{H^+}$ ratios significantly decreased 
in their passage to the WIM?  This is obviously a function of how clumpy and dusty 
the ISM is in the neighborhood of massive star formation regions.  Although these
two are important issues, the main motivation of this investigation is to determine if the 
$n_{He^+}/n_{H^+}$ ratios are systematically decreased in ionized gas surrounding 
the earliest stages of massive star-forming regions. 

Lyman continuum radiation leaking out of \HII\ regions ionizes
gas in its immediate vicinity as well as at large distances from the
\HII\ region. The 
collection of ionized gas surrounding compact \HII\ regions, referred to as
envelopes of \HII\ regions, in the inner Galaxy form the DIR \citep{a86}. 
The hierarchical structures in molecular cloud/ISM produce similar morphology
at different physical scales and at early stages of star-formation. 
For example, many Ultra-Compact \HII\ (\UCHII\ ) 
regions are known to have extended, diffuse ionized gas -- referred to as
envelopes of \UCHII\ regions or \UCHII\ envelopes -- associated with them (
\nocite{garayetal93}Garay \etal\ 1993, \nocite{kurtzetal99}Kurtz \etal\ 1999, 
\nocite{kimkoo01}Kim \& Koo 2001; see also \nocite{church02}Churchwell 2002). 
\UCHII\ regions are ionized by massive stars that are still embedded
in their natal molecular cloud thus representing a very early stage of star-formation.  
The morphology of \UCHII\ regions and their envelopes is 
similar to that of compact \HII\ regions and the DIR. The emission
measure of these envelopes is typically \gsim~a few times $10^3$ pc cm$^{-6}$, an
order of magnitude larger than the emission measure of DIR. The 
envelops of \UCHII\ regions absorb more than 65\% of the
ionizing radiation from the embedded stars \citep{kimkoo01}, which is
comparable to the percentage of UV photons absorbed by DIR.  
Thus observing He RRLs from higher emission measure diffuse gas surrounding
\UCHII\ regions may provide a clue to resolve the He ionization problem in the DIR. 
In Section~\ref{sec:ssource}
we describe the selection of sources for observation and discuss their properties.
The observations and data analysis are discussed in Section~\ref{sec:obs}.
Our main results are presented in Section~\ref{sec:res} and a discussion
of the results are given in Section~\ref{sec:dis}. 
A summary of the main results is given in Section~\ref{sec:sum}.
Appendix~\ref{a1} to \ref{a3} gives the details of the analysis discussed
in Section~\ref{sec:dis}. 

\section{Source Selection}
\label{sec:ssource}

A systematic study of continuum and RRL emission toward 
16 \UCHII\ regions was done by \citet{kimkoo01}. They detected diffuse,
extended emission toward 14 \UCHII\ regions in their sample. The similarity of
the LSR velocity of hydrogen RRLs from the \UCHII\ regions and
the diffuse gas suggests that the \HII\ region and the diffuse component 
are associated (see Section~\ref{lsrvel} for further discussion on
velocity structure based on our data set). This association is also suggested by 
the continuum morphology of these sources. 
The continuum emission from \UCHII\ regions and 
their envelope was used to estimate the Lyman continuum photon flux required
to maintain ionization. This Lyman continuum photon flux is used
to estimate the required stellar spectral type.  The spectral types for
the 14 \UCHII\ regions range from O4 to O9. We selected 
three sources (G10.15-0.34, G23.46-0.20 \& G29.96-0.02)
with embedded stars of type O5.5 or earlier for helium
RRL observations. \HII\ region models with
a single ionizing star suggest that
the helium and hydrogen Str\"{o}mgren
spheres will overlap for these sources. Hence
helium is expected to be singly ionized in the envelope
of the selected \UCHII\ regions. 

VLA 21 cm images of the three selected sources
G10.15-0.34, G23.46-0.20 \& G29.96-0.02
 are shown in Figs.~\ref{fig1},
~\ref{fig2} and ~\ref{fig3} respectively. These images are from the data obtained by
\citet{kimkoo01} and have an angular resolution of $\sim$ 40\arcsec $\times$ 20\arcsec. 
We show {\it Spitzer} three-color images of the three targets 
in Fig.~\ref{fig4a},\ref{fig4b} \& \ref{fig4c}, with GLIMPSE 3.6 and 8.0\,$\mu$m data in blue and green 
\citep{benjamin03, churchwell09} and MIPSGAL 24\,$\mu$m data in red \citep{carey09}.  
The red MIPSGAL emission is from warm dust grains spatially coincident with 
the ionized gas in \HII\ regions.  
The green GLIMPSE 8.0\,$\mu$m emission is dominated by polycyclic aromatic hydrocarbons (PAH) in the 
photodissociation regions (PDRs). A brief description
of the selected sources are given below and a summary of their properties
is given in Table~\ref{tab1}. 

\subsection{G10.15-0.34}

The G10.15-0.34 region is part of the W31 star-forming complex
\citep{sg70}. Numerous \HII\ regions and star clusters are present in this
well known star-forming region \citep{beutheretal11}. G10.15-0.34 is one of the dominant
infrared and radio continuum sources in W31 and is referred to as W31-South. The \UCHII\ 
region G10.15-0.34 is located in the complex 
W31-South \citep{wc89}. The distance to W31 is very uncertain since different
indicators provide different distances ranging from near ($\sim$ 2 kpc) to far
($\sim$ 15 kpc) kinematic distances \citep{deharvengetal15}. 
Infra-red spectrophotometric analysis of O stars 
in the \HII\ region indicates that G10.15-0.34 is at 3.55 kpc \citep{moisesetal11,blumetal01}.
\HI\ absorption studies resolve the kinematic distance ambiguity
indicating a near distance of about 3.55 kpc 
\citep{uetal12}. Trigonometric parallax measurements provide a distance
to W31 complex of 4.95 kpc \citep{sannaetal14}. The measured parallax  
is for the H$_2$O maser source in G10.62-00.38, one of the \HII\ regions in the
W31 complex.  This \HII\ region is about 0\ddeg.5 away from G10.15-0.34,
which corresponds to a projected separation on the sky of about 30 to 40 pc 
depending on the assumed distance. 
It is possible that the two \HII\ regions are at different distances. 
Here we adopt a distance of 3.55 kpc for the G10.15-0.34 region 
and for the \UCHII\ region G10.15-0.34 (see Table~\ref{tab1}).

The \UCHII\ region G10.15-0.34 is located at the western peak in the 21 cm 
image of the region G10.15-0.34 (see Fig.~\ref{fig1}, \nocite{kimkoo01}Kim \& Koo 2001). The
diffuse gas surrounding the \UCHII\ region has an angular extent  of
10\arcmin.9 $\times$ 6\arcmin.7, which corresponds to a linear
size of 11.3 pc $\times$ 6.9 pc at the distance of 3.55 kpc.  
The Lyman continuum photon flux (not corrected for dust extinction) 
obtained from the 21 cm continuum
flux density for the diffuse gas is 3.5 $\times 10^{49}$ s$^{-1}$,
corresponding to a single ionizing star of type O3 V \citep{msh05}. We use
a temperature for the ionized gas of 8000 K and a flux density at 21 cm  
of 55.22 Jy \citep{kimkoo01} to estimate 
the Lyman continuum photon flux \citep{rubin68}. The Lyman continuum
photon flux obtained from 21 cm emission is a lower limit
of the ionizing luminosity due to radio continuum optical depth effects
and possible escape of UV photons. 
The diffuse 21cm emission spatially coincides 
with a complex ionization ridge seen in the
high angular resolution (7\arcsec.5  $\times$ 4\arcsec.3)
5 GHz continuum image of G10.15-0.34 \citep{ghoshetal89}. 
NIR spectrophotometric study of a 1.\arcmin7 $\times$ 1.\arcmin8
region centered at RA 18:09:26.71, DEC $-$20:19:29.7 (J2000;
roughly coincides with the position G10.15-0.34a in Table~\ref{tab2}) 
has identified four O5.5-type stars, which together can account for
most of the radio derived Lyman continuum emission of the \UCHII\ envelope
\citep{blumetal01}. 

\subsection{G23.46-0.20}

The 21 cm image of the G23.46-0.20 region is shown in Fig~\ref{fig2}
\citep{kimkoo01}. The \UCHII\ region G23.455-0.201
\citep{wc89} is located slightly north of the strongest continuum peak 
in the 21 cm emission (marked as a star in Fig~\ref{fig2}).
The LSR velocity of H76$\alpha$ RRL
observed toward the \UCHII\ region is 99.0 \kms \citep{kimkoo01}. 
\citet{seiloetal04} analyzed the LSR velocity of the molecular cloud
associated with the \UCHII\ region and placed it at the near
kinematic distance of about 6 kpc (see \nocite{wienenetal15}Wienen et al. 2015). 
The distance to 
the 12 GHz methanol source G23.44-0.18 obtained from parallax measurements is 
5.88 kpc \citep{brunetal09}. Based
on the parallax measurement and LSR velocity study we adopt the
distance to the G23.46-0.20 region as 6 kpc (see Table~\ref{tab1}). 

The G23.46-0.20 region is located in the direction of the Galaxy
where several stellar clusters, supernova remnants (SNR) and giant molecular
clouds are present \citep{messineoetal14}. The diffuse gas 
surrounding the \UCHII\ region exhibits two peaks in the 21 cm 
continuum emission separated in the north-south direction (see Fig.~\ref{fig2}). 
The strongest peak (marked as a triangle in Fig.~\ref{fig2}) coincides with 
the IR source G23.437-0.209 \citep{cconti03}. The 21 cm continuum
emission is elongated in the east-west direction, part of which coincides
with the shell of the W41 SNR \citep{lt08}. The angular 
size of the G23.46-0.20 region is 8\arcmin.8 $\times$ 5\arcmin.8. At the distance
of 6 kpc, the linear size of the diffuse emission region is 15.4 pc $\times$ 10.1 pc. 
The Lyman continuum luminosity (not corrected for dust extinction
and contamination from W41 SNR emission)
obtained from the integrated radio flux density at 21 cm 
(11.31 Jy; \nocite{kimkoo01}Kim \& Koo 2001) 
is 4.0 $\times 10^{49}$ s$^{-1}$, for an assumed ionized gas temperature of 8000 K.
If a single star is responsible for the ionizing luminosity, then
the type of the star is O4.5 V \citep{msh05}.

\subsection{G29.96-0.02}

G29.96-0.02 is a well known star-forming region in the inner
Galaxy. It is part of the W43 complex and is referred 
to as W43-South. The 21 cm continuum image of the region
is shown in Fig.~\ref{fig3} \citep{kimkoo01}. The location of
the cometary \UCHII\ region G29.95-0.01 \citep{wc89} coincides with the 
strongest continuum peak (see Fig.~\ref{fig3}). Trigonometric
parallax measurements were made toward two 12 GHz methanol sources in
the G29.96-0.02 region -- one associated with the \UCHII\ region G29.95-0.01 
and the second towards G29.86-0.04. The distance obtained to the two
sources are 6.21 kpc and 5.26 kpc respectively \citep{zhangetal14}. 
Here we adopt 6.2 kpc as the distance to the G29.96-0.02 region (see Table~\ref{tab1}). 

The 21 cm emission from G29.96-0.02 is extended over 6\arcmin.3 $\times$ 5\arcmin.2,
corresponding to a linear size of 11.7 pc $\times$ 9.4 pc. 
The NRAO/VLA Sky Survey (NVSS), which has an angular resolution of 45\arcsec, 
has cataloged 10 sources consisting of both unresolved and extended objects \citep{cetal98}.
The brightest NVSS source is associated with the cometary \UCHII\ region
G29.95-0.01 \citep{wc89}. The physical properties of the NVSS sources
were obtained from their radio continuum emission \citep{beltranetal13}. 
These sources harbor stars of type O5 to B0.
The giant \HII\ region G29.944-0.042 is located south-east of G29.95-0.01. 
The Lyman continuum photon flux estimated using the 
flux density at 21 cm (12.69 Jy; \nocite{kimkoo01}Kim \& Koo 2001) is 4.8 $\times 10^{49}$ s$^{-1}$.
The temperature of the ionized gas is assumed to be 8000 K for estimating the
Lyman continuum luminosity and no dust extinction 
correction is applied. The luminosity can be produced by a single ionizing
star of type O4 V \citep{msh05}. K band spectroscopy
shows that the central exciting star of the \UCHII\ region
G29.95-0.01 is of type O5 \citep{hansonetal05}. 
Note that the estimated Lyman continuum luminosity 
is only 44\% of that obtained
for the giant \HII\ region G29.944-0.042 from 6 cm continuum observations
\citep{contic04}. The 6 cm source size obtained for G29.944-0.042 is
3\arcmin.7 \citep{kc97}. We attribute this discrepancy to radio continuum
optical depth effects at 21 cm. 

\section{Observation and Data Reduction}
\label{sec:obs}

\begin{figure}
\includegraphics[height=6.5in, width=7.0in, angle=0]{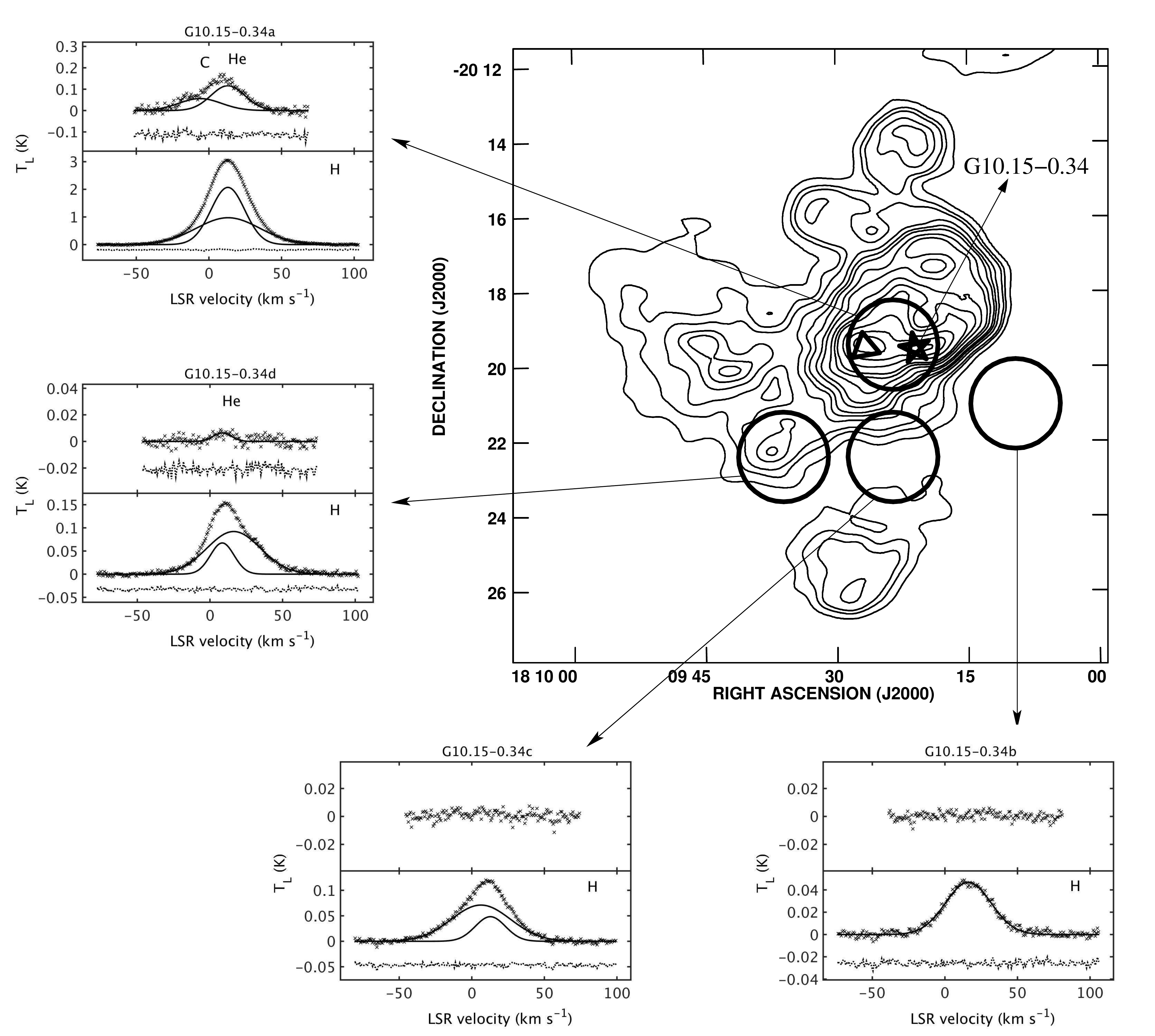}
\caption{
21cm continuum image \citep{kimkoo01} of G10.15-0.34 region is shown on top-right. 
The contour levels are (1,3,5,10,15,20,30,40,60,80,100,150,200,250,300)$\times 10^{-2}$ Jy/beam.
The observed positions are indicated by the continuum image with circles of radius the size 
of the GBT beam (2\arcmin.5). The observed spectra (indicated in 'x'), the Gaussian
line components (continuous line) and the residual spectra (dotted line) are also shown in the figure. The
position of the \UCHII\ region G10.15-0.34 is indicated by star. The center of NIR
observing region which has identified four O5.5-type stars (see text) is indicated by triangle.
\label{fig1} }
\end{figure}

\begin{figure}
\includegraphics[height=6.5in, width=7.0in, angle=0]{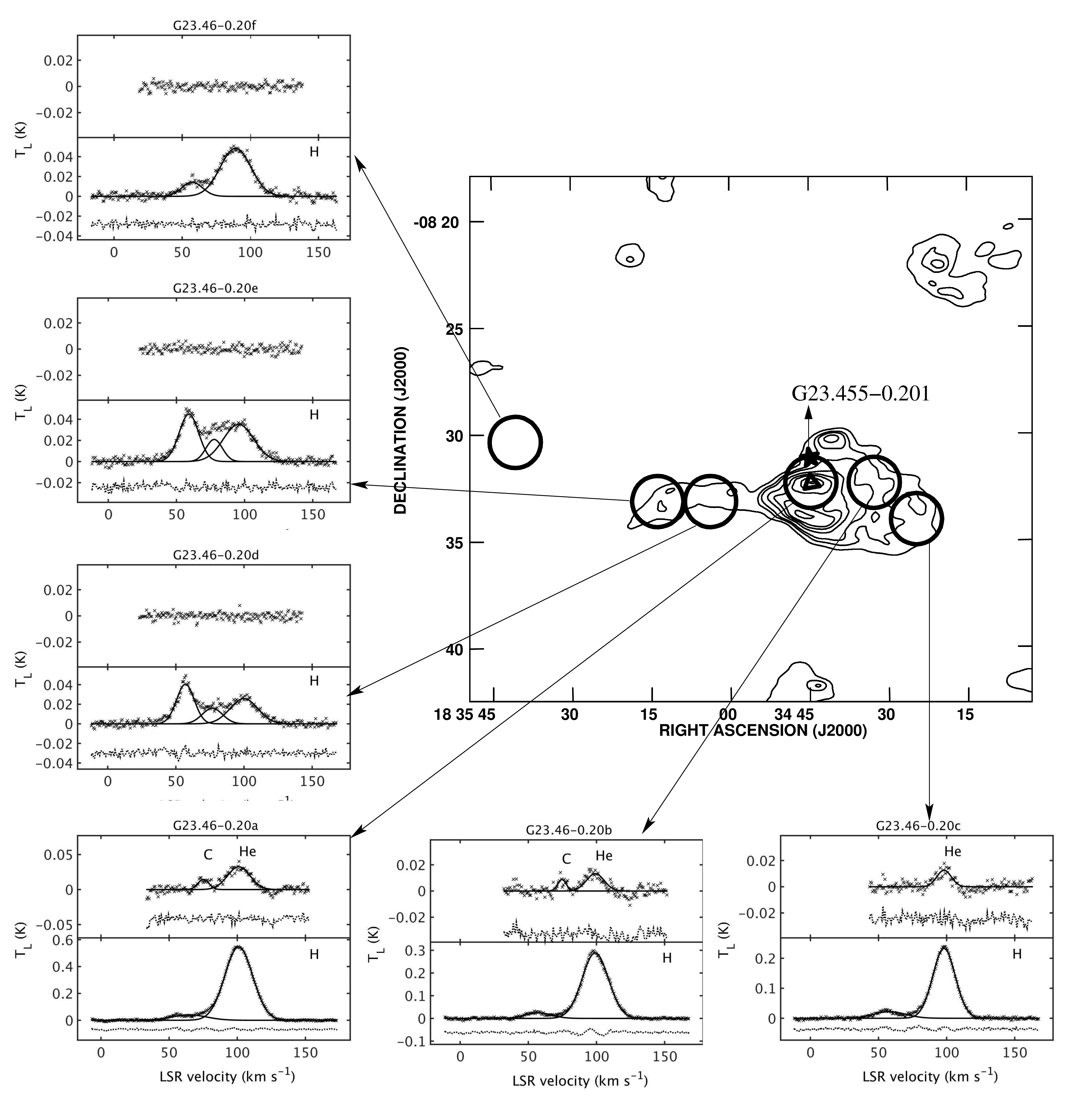}
\caption{
Same as Fig.~\ref{fig1} except for the source G23.46-0.20. The position
of the \UCHII\ region G23.455-0.201 is indicated by star. The position of
the IR source G23.437-0.209 (see text) is indicated by triangle. 
\label{fig2} }
\end{figure}

\begin{figure}
\includegraphics[height=6.5in, width=7.0in, angle=0]{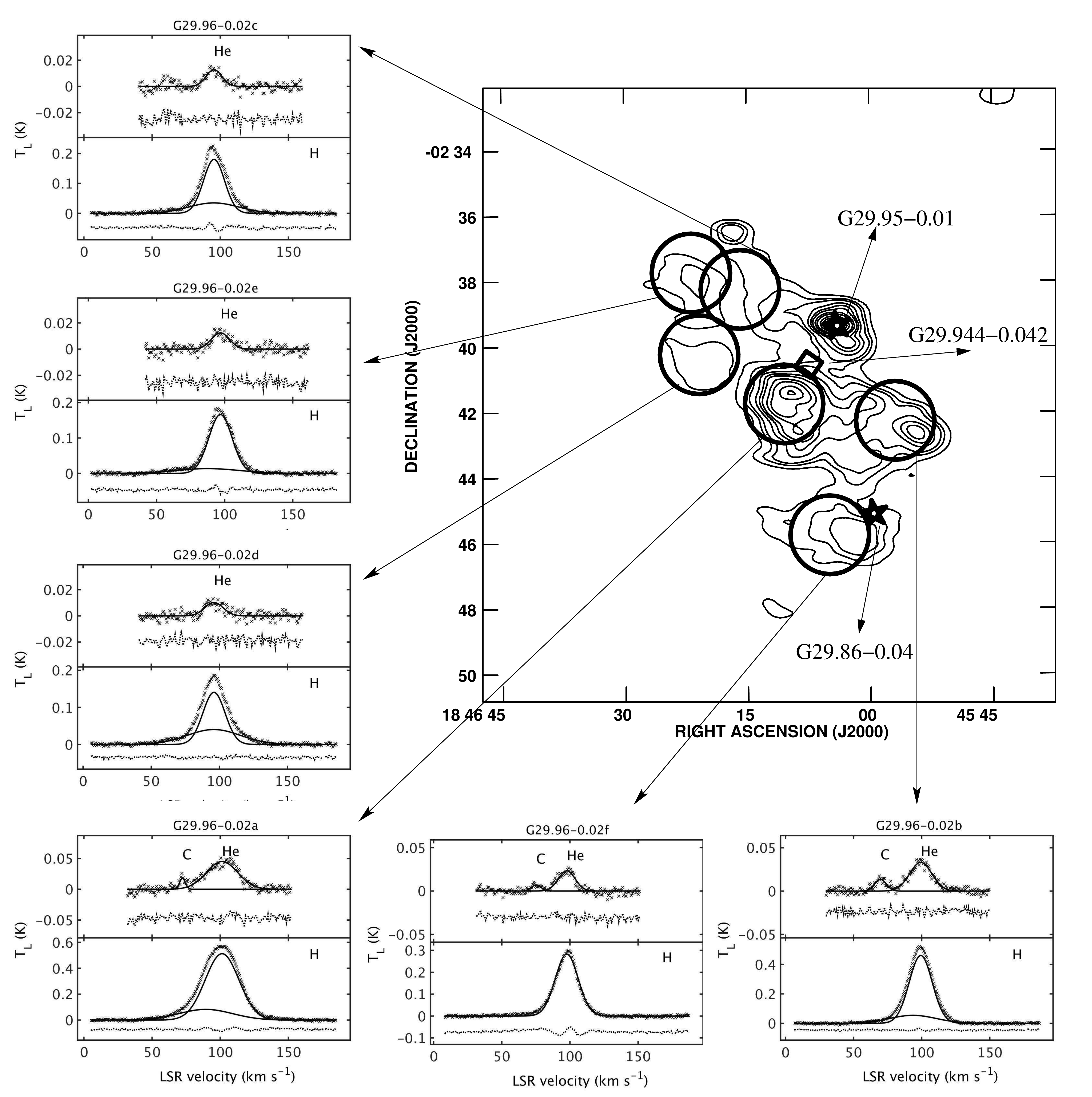}
\caption{
Same as Fig.~\ref{fig1} except for the source G29.96-0.02. The position
of the \UCHII\ region G29.95-0.01 and 12 GHz methanol source G29.86-0.04
are indicated by star and the position of the giant \HII\ region G29.944-0.042
is indicated by diamond.
\label{fig3} }
\end{figure}

\begin{figure}
\includegraphics[width=6.0in, angle=0]{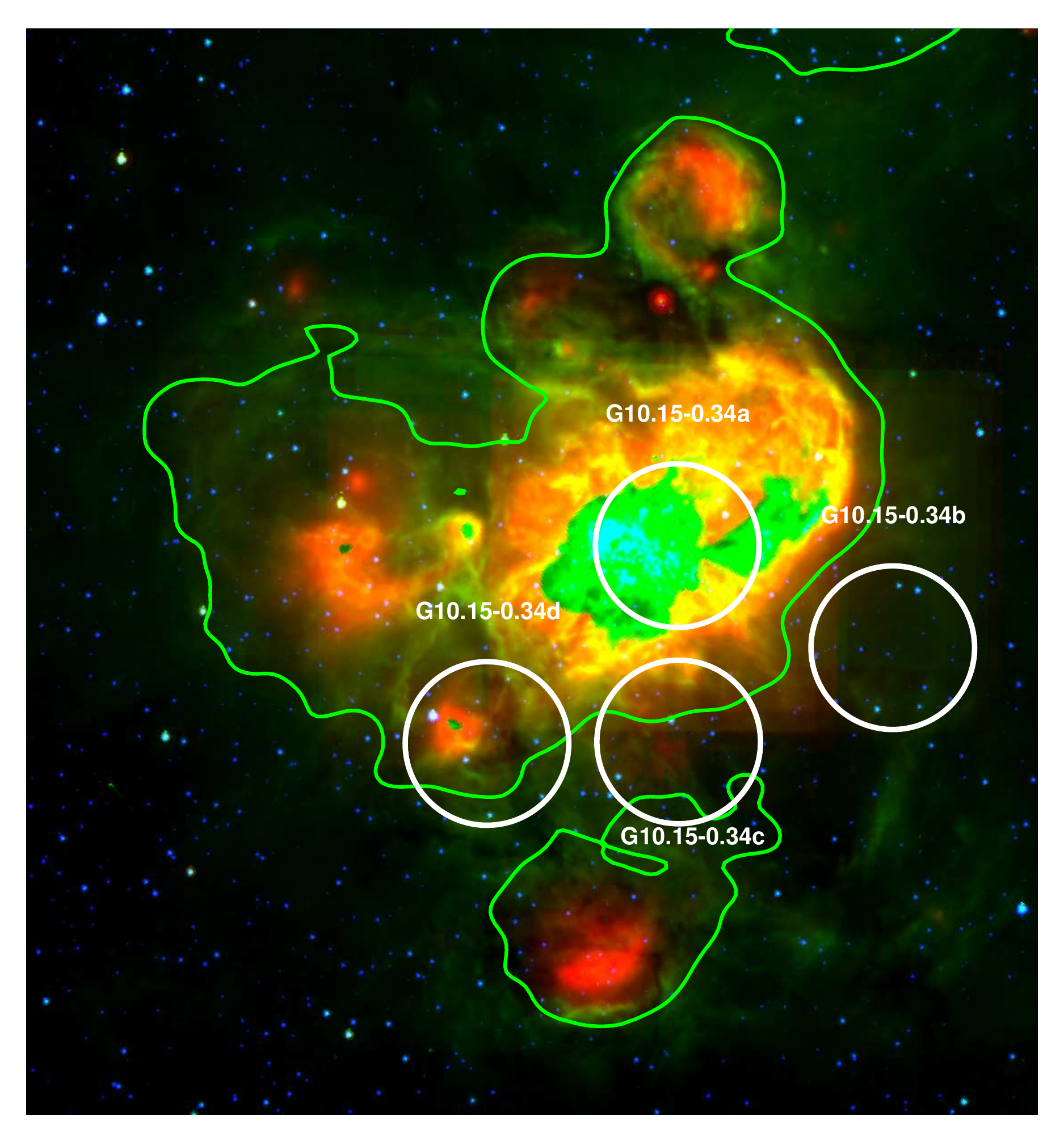}
\caption{
{\it Spitzer} three-color images of the \UCHII\ envelope G10.15-0.34. 
GLIMPSE 3.6 and 8.0\,$\mu$m data are shown in blue and green
\citep{benjamin03, churchwell09} and MIPSGAL 24\,$\mu$m data in red \citep{carey09}.
The observed positions are indicated in circles of diameter equal to the size of 
the GBT beam. The image spans the same sky region as shown in Fig.~\ref{fig1}. 
Note that near the position G10.14-0.34a the measured flux density at 
24\,$\mu$m is affected due to detector saturation. The 21 cm
radio contour of level $10^{-2}$ Jy/beam is also shown.
\label{fig4a} }
\end{figure}

\begin{figure}
\includegraphics[height=6.0in,  angle=0]{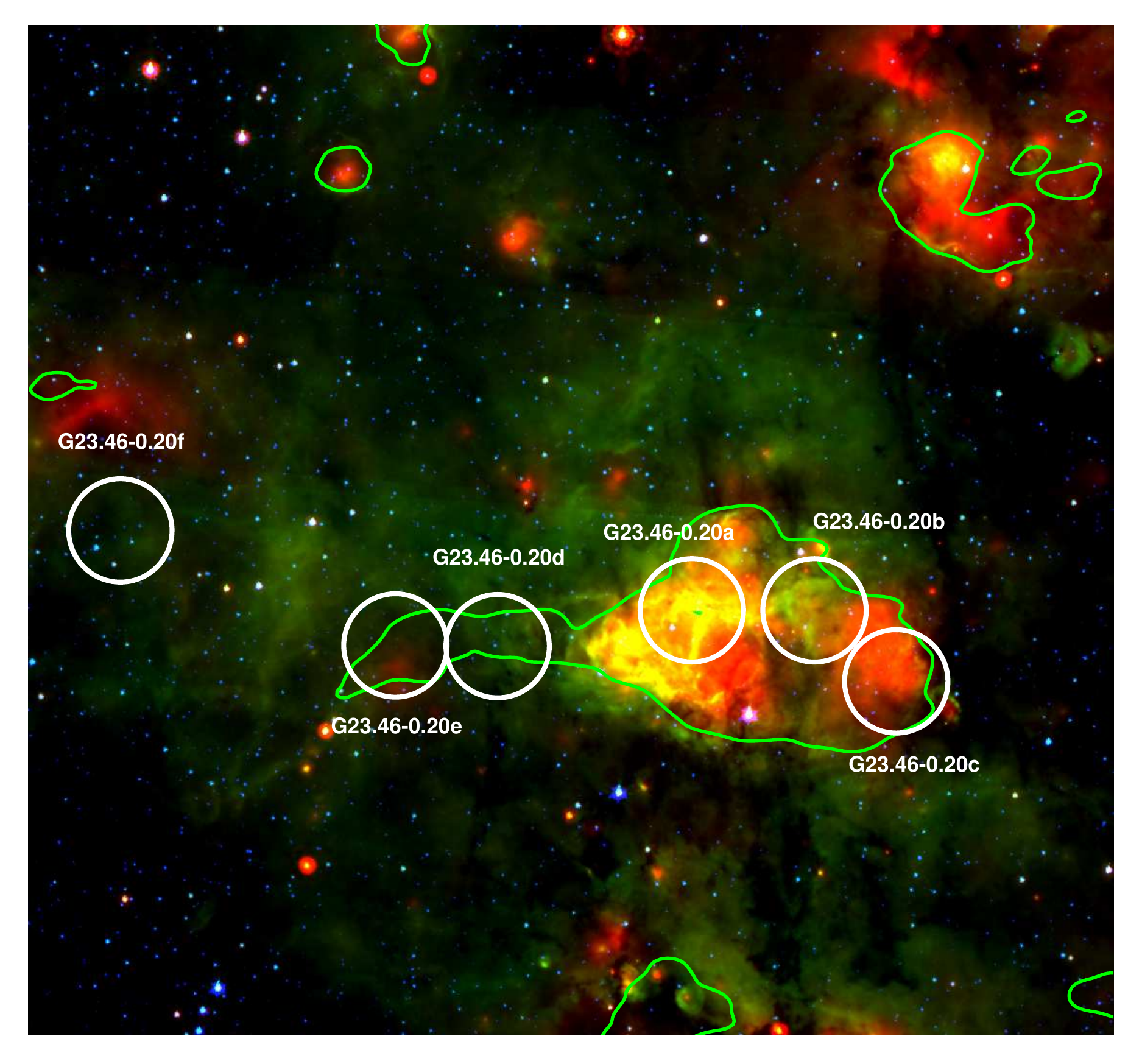}
\caption{
Same as Fig~\ref{fig4a} except for the \UCHII\ envelope G23.46-0.20.
The image spans the same sky region as shown in Fig.~\ref{fig2}. 
\label{fig4b} }
\end{figure}

\begin{figure}
\includegraphics[height=6.0in, angle=0]{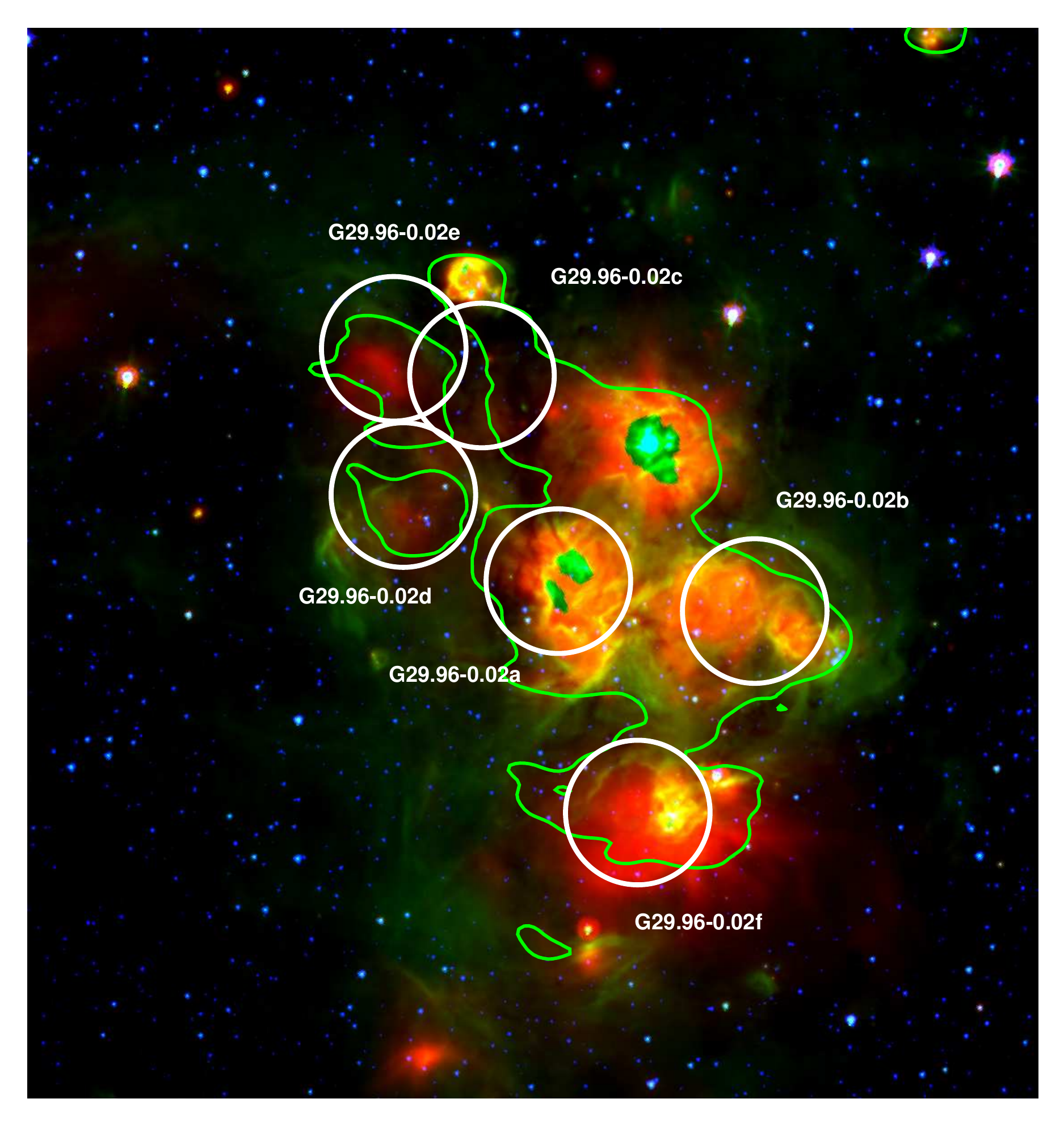}
\caption{
Same as Fig~\ref{fig4a} except for the \UCHII\ envelope G29.96-0.02.
The image spans the same sky region as shown in Fig.~\ref{fig3}. 
Note that near the position G29.96-0.02a the measured flux density at 
24\,$\mu$m is affected due to detector saturation.
\label{fig4c} }
\end{figure}

The sizes of the selected sources are listed
in Table~\ref{tab1}. The typical physical size of the
selected diffuse regions surrounding \UCHII\ regions is \gsim 9 pc, which 
corresponds to an angular size \gsim 5$^{'}$. We made 
RRL observations toward 13 positions in the selected sources with 
the Robert C. Byrd Green Bank Telescope (GBT) near 5 GHz. 
The FWHM (full width at half power) beam width of the telescope
near 5 GHz was about 2\arcmin.5, allowing us to sample physical
scales smaller than the sizes of the diffuse regions
in the selected sources.
The J2000 RA and DEC of the observed positions are listed in Table~\ref{tab2}
and are shown in Figs.~\ref{fig1},\ref{fig2} \& \ref{fig3}.
For comparison, we also observed 3 positions (G10.15-0.34a, 
G23.46-0.20a, G29.96-0.02a;
see Figs.~\ref{fig1},\ref{fig2} \& \ref{fig3} and Table~~\ref{tab2}) 
toward peaks in the 21 cm continuum images of the selected regions. 

We simultaneously observed eight RRL transitions (104$\alpha$, 105$\alpha$, 106$\alpha$, 109$\alpha$,
110$\alpha$, 111$\alpha$, 112$\alpha$ \& 113$\alpha$)
of hydrogen, helium and heavy elements. The reference spectra to correct for bandpass shape 
were obtained by switching the frequency by 8.5 MHz ($\sim$ 530 \kms).  
GBTIDL routines were used to correct for bandpass shape for each RRL transition 
and calibrate the spectra in units of antenna temperature.
Doppler tracking was done using the H110$\alpha$ RRL transition
during observation. The residual Doppler correction in other RRL 
transitions was done offline by shifting them in LSR velocity. 
The Doppler corrected spectra for each RRL transition were
averaged using GBTIDL routines. 
The rest of the data analysis was done using
routines developed in Matlab. The average spectrum for each transition
was examined for radio frequency interference (RFI).
After editing RFI affected spectra and removing spectra that have 
beta transitions contaminating the helium line, we 
re-sampled the sub set of RRL spectra to a common velocity resolution.
This sub set was averaged to obtain the final integrated spectrum. The
RRL transitions averaged to get the final spectrum 
are listed in Table~\ref{tab2}. The actual
observing time and effective integration time, obtained from the
number of transitions averaged, for each position is also included
in Table~\ref{tab2}.
The effective integration time is the actual integration
time of the final spectra after editing out the RFI affected spectra
and averaging the data corresponding to the listed transitions in 
Table~\ref{tab2}.  
A 4$^{th}$ order polynomial was subtracted from the final spectrum.

\section{Results} 
\label{sec:res}

Results of Gaussian component analysis of the final spectra 
are included in Table~\ref{tab3}. The source name, peak line 
amplitude in K, FWHM line width
in \kms, LSR velocity of the line in \kms\ and the atom 
responsible for the line emission
are included in Table~\ref{tab3}.  The final spectra
and the Gaussian components are shown in Figs.~\ref{fig1}, ~\ref{fig2}
and ~\ref{fig3}. The signal-to-noise ratio of the hydrogen line is greater
that 10 $\sigma$ in most cases allowing us to fit multiple Gaussian components
to the line profile. We choose the minimum number of Gaussian components 
for the hydrogen line so that the residuals after subtracting the Gaussian model are 
consistent with the RMS of the spectral noise. Since we did not know a 
priori which hydrogen line component
corresponds to the helium line, we tried several methods to obtain 
the helium line parameters. We found that the different methods provide
a slightly different value for the $\frac{n_{He^+}}{n_{H^+}}$ ratio (see below), 
but all these values are consistent within 1.5$\sigma$ estimation error of this ratio. 
We finally adopted the following strategy to get the helium line parameters --  
we set the LSR velocity of the helium line equal to the central velocity
of the strongest hydrogen line while fitting the Gaussian components.
The exception is for the position G10.15-0.34d where the helium line
velocity is close to the 8.45 \kms\ hydrogen line component and hence
we used this velocity for the central velocity of helium. No constraints
were used to obtain line parameters of atoms heavier that helium.
The heavy element
is tentatively identified as carbon based on its frequency and its relatively high
abundance compared to other heavy elements with 
ionization potential $<$ 13.6 eV. 

\subsection{LSR velocity of hydrogen line}
\label{lsrvel}

The LSR velocity structure of the hydrogen lines toward the observed sources was
studied earlier by \cite{kimkoo01}. We re-examine the line velocity structure
with our new, sensitive RRL observations. The line structure in general is
complex, exhibiting multiple velocity components toward all the three sources. 
Toward the 4 observed positions in G10.15-034, the peak hydrogen line velocity 
ranges between 16.3 and 6.3 \kms, with a mean value of 12.3 \kms. The
velocity spread of this source is attributed to bipolar-flow exhibited
by the ionized gas \citep{kimkoo01}. A bipolar-flow is also inferred from the velocity
structure of dense molecular tracers observed toward G10.15-034 
\citep{kimkoo03, deharvengetal15}. The mean velocity of the molecular tracers
is in good agreement with that of the RRLs. Thus the diffuse gas observed 
toward G10.15-034 is likely to be associated with the \UCHII\ region G10.15-034 
as concluded earlier by \cite{kimkoo01}. The size of the diffuse region is
about 8 pc (see Table~\ref{tab1}), much larger than the \UCHII\ region, and
hence multiple massive stars in the star-forming region will be contributing 
to the ionization of the diffuse gas. 

The hydrogen line toward G23.46-0.20 shows 
at least three LSR velocity components -- $\sim$ 60, 76 and 95 \kms. 
The LSR velocity of hydrogen line 
observed toward the \UCHII\ region is 99 \kms\ \citep{kimkoo01} and
hence it is likely that only the 95 \kms\ component is associated with the source 
G23.46-0.20. This is because, there is no evidence for outflows with such large
velocity difference to produce the other line components. Further, the kinematic
distances of the 60 and 76 \kms\ line components are greater than a kpc
from that of the 95 \kms\ component. 
Thus the 60 and 76 \kms\ components are most likely due to ionized gas 
along the line of sight.  The 95 \kms\ component in our data shows
a velocity range from 89.2 to 101.1 \kms\ over the 6 observed positions
toward the source, with a mean value of 97.2 \kms.
The $^{13}$CO line observed toward G23.46-0.20 shows a similar 
velocity range. The velocity structures of molecular line and RRL 
have been interpreted as due to a champagne flow \citep{kimkoo03}
and not due to multiple ionized gas location along the line-of-sight. 
As in the case of G10.15-034, the large size
of the diffuse gas (12 pc; see Table~\ref{tab1}) implies multiple 
stars present in the star-forming region G23.46-0.20 may be 
contributing to the ionization of the diffuse region. 

The LSR velocity range of hydrogen line observed toward 6 positions
in G29.96-0.02 is between 87.3 and 101.4 \kms, with a mean value of 95.9 \kms. The
mean velocity is similar to the LSR velocity of RRLs (95.3 \kms) observed
toward the \UCHII\ region \citep{wc89}. Molecular line observations show
a velocity gradient from $\sim$ 92 to 100 \kms\ roughly in
agreement with that observed in RRLs. There are several 
compact \HII\ regions embedded in the diffuse ionized gas.
Thus the velocity structure of the diffuse gas is likely to be
reflecting the blending of ionized gas from the different
compact sources.     

To summarize, in all three observed sources, we could identify
a line component that has LSR velocity similar to that of the
\UCHII\ regions. We consider that this line component is 
due to emission from the diffuse ionized
gas surrounding the \UCHII\ regions. In the following sections,
we refer to the gas with the identified velocity component as the 
\UCHII\ envelope. It is likely that
several ionizing sources embedded in the diffuse gas 
are contributing to the ionization of this gas.  

\subsection{Helium to hydrogen line ratio}

The ratio 
\be
\frac{n_{He^+}}{n_{H^+}} = \frac{\int T_{L,He} dv}{\int T_{L,H} dv},
\label{lineanttemp}
\ee
obtained from the RRL data is given in Table~\ref{tab4}. Here $T_{L,He}$ and 
$T_{L,H}$ are the line antenna temperatures of helium and hydrogen, respectively (see
Table~\ref{tab3}), and the integration is over LSR velocity. 
The computed value of $n_{He^+}/n_{H^+}$ ratio along with 1 $\sigma$ error as well as upper 
limits are listed in column 2 in Table~\ref{tab4}. The velocities
of the hydrogen line components used to obtain the ratio are given in 
column 3.
In cases where helium line is not detected, the upper limit for the ratio is computed
using the 1$\sigma$ value at the line
free region of the spectrum and the net line width of the hydrogen line components
listed in column 3 as the expected width of the helium line.
The other columns are source name
(column 1), offset distance from the `ionization center' (column 4; see below),
4.875 GHz brightness temperature (column 5; see below) 
and emission measure (column 6). The emission measures are obtained from
the main beam brightness temperature, estimated from the 4.875 GHz continuum survey of 
\citet{aetal79}, which has similar angular resolution as that of the GBT observations.
The emission measure is obtained using the equation \citep{mh67}
\be
EM = \frac{T_{b4.9}}{8.235 \times 10^{-2} T_e^{-0.35} \nu^{-2.1}}
\ee 
where $EM$ has units pc cm$^{-6}$, $T_{b4.9}$ is the brightness temperature
in K at the frequency $\nu = 4.875$ GHz,
$T_e$ is the electron temperature in K (see Table~\ref{tab1}). 

The helium lines are detected toward G10.15-0.34a \& d in G10.15-0.34. 
We use the 24 $\micron$ emission in the MIPSGAL 
data, which has higher angular resolution ($\sim$ 6\arcsec) than the 21 cm images,
to identify compact \HII\ regions (emission shown in red in Fig.~\ref{fig4a}) in the 
observed positions. This identification can be done because, a large part of 
the 24 $\micron$ emission originates from very small dust grains
(few nanometer in size) inside \HII\ regions heated by UV radiation
from the star and collision with ionized gas particles 
\citep{pavlyuchenkovetal13}. The observed
correlation between the 24 $\micron$ emission and 21 cm flux density
of Galactic \HII\ regions supports this picture \citep{andersonetal14}. 
Both these positions have bright 24 $\micron$ emission
indicating the presence of compact \HII\ regions. Helium lines are
not detected toward G10.15-0.34c \& b. No compact \HII\ regions are
present in these directions as inferred from the 24 $\micron$ image. 
We averaged the GBT spectra observed toward these two positions
to get a stringent upper limit on helium line emission. The line
parameters obtained from the average spectrum is listed in Table~\ref{tab3}.
The LSR velocity of the hydrogen line (12.8 \kms) is similar to those
observed toward other positions in the source G10.15-0.34. 
Thus the positions G10.15-0.34c \& b sample the more diffuse region
of the \UCHII\ envelope. The 1$\sigma$ upper limit for helium line 
emission and the value of the $\frac{n_{He^+}}{n_{H^+}}$ toward 
the diffuse region are $\sim$ 2 mK and 0.03 respectively, both obtained
from the average spectrum.   

Helium lines are detected toward positions G23.46-0.20a,b \&c.
Examination of the 24 $\micron$ image indicates that compact
\HII\ regions are present in these regions (see Fig.~\ref{fig4b}). No helium lines were
detected toward positions G23.46-0.20d,e \&f. Toward the position
G23.46-0.20e a compact \HII\ region is present as inferred from
its 24 $\micron$ emission. The hydrogen lines 
toward positions G23.46-0.20d \&f have components with LSR
velocities 99.9 and 89.2 \kms~respectively. The strongest hydrogen line 
components toward positions G23.46-0.20a,b,c \& e shows a gradient 
in LSR velocity with values reducing from  101.1 \kms~to 95.6 \kms.
Thus the diffuse ionized gas toward G23.46-0.20d as well as G23.46-0.20f 
may be associated with G23.46-0.20. We averaged the
GBT spectra toward G23.46-0.20d \&f to get a stringent upper limit
on helium line emission. The upper limit for the value of the $\frac{n_{He^+}}{n_{H^+}}$ 
obtained from this average spectrum is 0.05.
 
Helium lines are detected toward all the observed positions in
G29.96-0.02. All these positions have associated
bright 24 $\micron$ emission indicating the presence of compact
\HII\ regions (see Fig.~\ref{fig4c}). The value of the $\frac{n_{He^+}}{n_{H^+}}$ 
varies between 0.53 and 0.81 at the observed positions. Such
variation is also observed toward the sources G10.15-0.34
and G23.46-0.20 at positions where helium lines are detected. 

Fig.~\ref{fig6} shows the histogram of the values for $n_{He^+}/n_{H^+}$ ratio
obtained toward positions where helium lines are detected from all the
three sources. The values of the ratio from our sample 
range from 0.033 to 0.081; the mean value is 0.058. 
The mean value for $n_{He^+}/n_{H^+}$ ratio
is consistent with those observed toward compact \HII\ regions in the Galaxy
\citep{churchwelletal74, quirezaetal06}. It may be noted that some of the \HII\ regions  
observed by \citet{churchwelletal74} and \citet{quirezaetal06} may
be ionized by stars later than O6 and hence the mean value of $n_{He^+}/n_{H^+}$
from their data set is biased toward lower values. 

Multiple OB stars are responsible for the ionization of the observed sources. Due
to the complexity of the source, however, it is difficult to identify a region,
referred to as the `ionization center',
where most of the massive stars are located in each source. We
assigned the `ionization center' for the region G10.15-0.34 as the center of 
the NIR emission observed by \citet{blumetal01}(RA 18:09:26.71, DEC $-$20:19:29.7 J2000; see 
Section~\ref{sec:ssource}; shown in Fig.~\ref{fig1}), which has revealed
4 O type stars. The `ionization center' for the region G23.46-0.20
is taken as the \UCHII\ region G23.455-0.201 and that
for the region G29.96-0.02 is taken as the \UCHII\ region G29.95-0.01.
In Fig.~\ref{fig7}, we plot the values of the $n_{He^+}/n_{H^+}$ ratio against 
the estimated distance offset
(see Table~\ref{tab4}) of the observed positions from the `ionization center'. 
We also plot the measured $n_{He^+}/n_{H^+}$ values against the 
estimated emission measure (see Table~\ref{tab4}) in Fig.~\ref{fig8}. 
The upper limits obtained from the average spectra are included in the plot with 
filled triangles. 
No particular trends are evident in these plots but it is clear that the
helium is not uniformly ionized in the observed regions.

\section{Discussion}
\label{sec:dis}

We first examine the observed $n_{He^+}/n_{H^+}$ ratio toward the 
continuum bright regions (G10.15-0.34a, G23.46-0.20a \& G29.96-0.02a). The
first two positions encompass the \UCHII\ regions G10.15-0.34 and G23.455-0.201
respectively. The third position is observed near the giant \HII\ region
G29.944-0.042. The mean value of the $n_{He^+}/n_{H^+}$ ratio obtained from
the data toward the three positions is 0.062$\pm$0.02. This value is slightly lower 
than the mean value of $n_{He^+}/n_{H^+}$ ratio (0.08$\pm$0.02)observed towards \HII\ regions
ionized by stars of spectral type O6 or hotter star \citep{setal83}. The origin
of this lower $n_{He^+}/n_{H^+}$ value may be related to the
non-uniform helium ionization
observed in the selected sources.  
 
In the diffuse regions in the sources G10.15-0.34 and G23.46-0.20 less
than 33\% and 51 \%, respectively, of the helium is singly ionized.
The detection of helium lines toward high emission measure
regions in the envelopes implies stars of type O6 or earlier
are embedded in these regions.  If these stars are responsible 
for the ionization of the diffuse gas then helium is expected
to be fully ionized. Thus, the situation in the \UCHII\
envelopes is similar to the WIM and DIR -- i.e. stars of type O6 or earlier
are required for the ionization of WIM and DIR but helium in these regions 
is observed to be not fully ionized. Observations toward WIM and DIR
probe helium ionization at distances greater than a few tens of parsec
from the ionizing stars. Our observations, on the other hand, probe helium
ionization in diffuse regions at distances \lsim~10 pc from newly born 
massive stars.  
Below we investigate possible origins 
for the non-uniform helium ionization. 

As mentioned earlier, stars hotter than $\sim$ O6 are the likely primary
ionization sources of the DIR and WIM.  A ``hardening'' of the radiation 
field with distance from the ionizing source is expected since the
the ionization cross-sections for both H and He 
decrease with energy proportional to $(h\nu)^{-3}$. If that is the case, can 
helium be doubly ionized in regions where He$^{+}$ lines are not detected ? 
The velocity range of RRL spectra 
obtained in our observation include 167$\alpha$, 175$\alpha$ and 178$\alpha$
recombination line transitions of He$^{++}$. The velocity range
of 167$\alpha$ transition overlaps with the H105$\alpha$ line. 
The 178$\alpha$ transition is affected by a bad baseline. We therefore
examined the 175$\alpha$ transition of He$^{++}$ but failed to detect the line
from any of the observed positions. We further averaged the spectra from all
the observed positions after resampling and shifting them in velocity
but again failed to detect the He$^{++}$ line.
The 1$\sigma$ upper limit we obtained is 0.004 K. Thus we
conclude that helium is not doubly ionized at the observed positions.

The sizes of the helium and hydrogen ionization zones depend on the
number of `helium' ($q_{He}$; i.e. photon wavelength range 228 \AA~$< \lambda < 504$ \AA) 
and `hydrogen' ($q_{H}$; i.e. photon wavelength range 504 \AA ~ $< \lambda < 912$ \AA) 
Lyman photons available for ionization \citep{mathis71}. It is convenient to
characterize the Lyman photon spectrum by the ratio $\gamma = \frac{q_{He}}{q_{H}}$ \citep{mathis71}. 
To understand physically the dependence of the ionization zone size to $\gamma$,
to a first approximation, we assume (a) that all the $q_{He}$ photons ionize helium and (b) the photons due 
to the recombination of helium ions ionize hydrogen. The first
assumption is justified since the ionization cross section of helium 
for energies $\ge$ 24.6 eV is larger compared to that of hydrogen. The second assumption
follows from the energy level of helium and the probability of the
paths involved in the recombination process. With the above assumptions and 
keeping in mind that the helium recombination rate is $\sim$ 18\% of 
that of hydrogen in region where both atoms are (singly) ionized, the ionization
equilibrium implies that the size of helium ionization zone is smaller
than the hydrogen ionization zone if $\gamma$ \lsim  0.2 
(see \citet{d11} page 172 for further reading). A numerical solution
to this problem was presented by \cite{mathis71}, where he derived the 
relationship between $\gamma$ and the sizes of the ionization zones of the
two atoms (see Appendix~\ref{a3}). The observed \UCHII\ envelopes are ionized by multiple OB stars,
which may be members of a cluster. In the early stage of star formation, 
it is likely that the massive members of the cluster are embedded in (locally) 
ionization bounded regions and may not be contributing to the ionization
of the diffuse regions in the envelopes, which then results in a change in 
$\gamma$ of the radiation field. Here we ask, if the diffuse regions
in the \UCHII\ envelopes are ionized by the rest of the stars in the cluster, what 
upper mass of the cluster member can produce the observed $n_{He^+}/n_{H^+}$ ratio ? 

In Appendix~\ref{a1}, we compute $\gamma$
as a function of the upper cutoff mass for the cluster. 
The result is shown in Fig.~\ref{fig9}. The observed
$n_{He^+}/n_{H^+}$ ratios are used to obtain
$\gamma$ for the radiation field as described in
Appendix~\ref{a3} (no selective dust absorption in considered here, i.e. $a_0$ is
taken as unity). The measured upper limit to $n_{He^+}/n_{H^+}$ ratio for G10.15-0.34 and G23.46-0.2
are $\sim$ 0.03 and 0.05 respectively. The estimated $\gamma$ for the 
two sources are, respectively, 0.04 and 0.06. Thus, from Fig~\ref{fig9}, it follows
that the upper mass of the cluster member that ionizes these \UCHII\  envelopes
has to be \lsim 30 \Msun to explain the observed value of the $n_{He^+}/n_{H^+}$ ratio.
The Lyman continuum emission estimated for G10.15-0.34 and G23.46-0.2
is $\sim$ 5 $\times 10^{49}$ s$^{-1}$ (average value for these two sources 
from Table~\ref{tab1}). If all the cluster members are contributing to 
the ionization, then the mean ionizing flux per solar mass is 
6.1 $\times 10^{46}$ s$^{-1}$ \Msun$^{-1}$ (see Appendix~\ref{a1}) 
implying a cluster mass of  $\sim$ 800 \Msun. For this cluster mass,  
statistical uncertainties in stellar population of 
the cluster are expected to be large, especially at the high mass end. 
We note here that our analysis does not take into account this 
statistical uncertainty and thus the upper mass cutoff
estimated above can be somewhat uncertain. Further observations are needed to
include the statistical uncertainty in the analysis and will be presented elsewhere.

The infra-red (IR) continuum emission from \HII\ regions are reprocessed stellar
radiation from dust. The bolometric luminosity of \HII\ regions estimated
from the IR emission can thus be used to infer the stellar type.  
For several \HII\ regions, the estimated Lyman continuum  
luminosities, obtained from their radio free-free emission, exceed those 
expected from their inferred stellar type \citep{wc89,sanetal13}. It has been
suggested that this excess ionizing radiation is due to 
accreting low-mass stars present in the star-forming region \citep{smith14}. 
The `hot' spots on the stellar surface
in such accreting stars form the additional source for Lyman continuum emission. 
We investigate whether the presence of such accreting low-mass stars in our 
observed sources can contribute to the hydrogen Lyman photons thus modifying 
the net $\gamma$.  The details of the analysis is given in Appendix~\ref{a2}.
We find that the effective temperature of 
the `hot' spot is \gsim~35000 K during a considerable 
fraction of the accretion phase\footnote{
Since the hot spot effective temperature is in excess of 35000 K, 
detecting helium RRLs from individual \HII\ 
regions with Lyman excess is another test for the `cold' accretion, 
hot spot model.}. Thus the contribution from 
accreting low mass stars modifies the net $\gamma$ of the ionizing
radiation in such a way that the helium is expected to be ionized (see Fig.~\ref{fig9}).
We conclude that ionizing radiation from accreting low-mass stars
cannot be the cause of the low $n_{He^+}/n_{H^+}$ ratio observed toward the selected sources.

Finally, we investigate whether selective absorption of Lyman photons due to
dust in the \HII\ regions can produce the observed low $n_{He^+}/n_{H^+}$ ratio, 
as suggested earlier by \citet{msc74}. Appendix~\ref{a3} summarizes the analysis. 
In this process, the dust absorption cross section of helium Lyman photons is larger than
that of hydrogen Lyman photons, which changes $\gamma$ as the photons propagate
through the \HII\ region. This change in $\gamma$ in turn reduces the 
size of the helium ionization zone, thus lowering the observed $n_{He^+}/n_{H^+}$ ratio. 
To estimate the change in $\gamma$, we first express the dust absorption cross sections 
in terms of the extinction cross section at 13.6 eV 
provided by the \citet{wd01} dust model. The ratio of the helium to hydrogen Lyman photon
absorption cross sections, $a_0$, is then estimated using the observed 
$n_{He^+}/n_{H^+}$ ratio and taking the $\gamma$ due to the stellar cluster as 0.2 (
see Appendix~\ref{a3}).  
The results are given in Table~\ref{tab5}
for different \HII\ region filling factors and $f_{uv}$. $f_{uv}$ is a scaling factor 
that takes into account of deviations in the cross section at 13.6 eV from the \citet{wd01} 
value. The estimated $a_0$ values range from $\sim$ 1.8 to 4.4, 
which are closer to those obtained by \citet{ps78} but a factor $\sim$ 1.8 smaller
than the values obtained by \citet{msc74}. In the \cite{wd01} dust model, the extinction cross section 
increases from $\sim$ 2 $\times 10^{-21}$ cm$^2$/H-atom at 13.6 eV to $\sim$ 4 $\times 10^{-21}$ cm$^2$/H-atom
at $\sim$ 18 eV and then declines below $\sim$ 0.2 $\times 10^{-21}$ cm$^2$/H-atom at 24.6 eV.
The extinction and absorption cross sections are related through the dust albedo (see
Appendix~\ref{a3}), which is not known in the relevant photon energy range.
The estimated values of $a_0$ indicates that the absorption cross section needs to be larger
by a factor $\ge$ 1.8 for photons of energy 24.6 eV compared to that at 13.6 eV. 
The dust model of \citet{wd01} is derived from extinction measurements toward 
diffuse ISM and consists of silicate and carbonaceous grains with size 4 $\times 10^{-4} \mu$m
to $\sim$ 0.5 $\mu$m. In \HII\ regions the grains with size \lsim  $10^{-3} \mu$m
(mostly PAH molecules) are destroyed in the central region 
close to the ionizing star \citep{pavlyuchenkovetal13}. The dominant
grains are the very small graphite grains, as inferred from the bright
24 $\mu$m emission from many \HII\ regions. Whether a self consistent 
modeling of dust emission from \HII\ regions can provide the above estimated 
properties for the absorption cross section of very small graphite grains 
needs further investigation.
Thus with the current data, we are unable to conclusively establish
whether selective absorption is the cause for low observed $n_{He^+}/n_{H^+}$ ratio. 

\section{Summary}
\label{sec:sum}

We observed helium and hydrogen RRLs near 5 GHz toward
envelopes of three \UCHII\ regions -- G10.15-0.34, 
G23.46-0.20 \& G29.96-0.02. This data set is used
to investigate helium ionization in the \UCHII\ envelopes. Our main
results are : 
\begin{enumerate}
\item
Our observations indicated that helium was not uniformly ionized
in the \UCHII\ envelopes. Toward G10.15-0.34 and G23.46-0.20
helium lines were not detected in a few positions and 
the upper limits obtained for the $n_{He^+}/n_{H^+}$ ratio 
were 0.033 and 0.051 respectively.
Our data thus show that helium is not fully ionized in the 
diffuse regions located at a few pc from the young massive stars 
embedded in the observed sources. 

\item
The mean value of the $n_{He^+}/n_{H^+}$ ratio obtained from the 
positions nearer to the peak of the radio continuum emission
in the observed sources was 0.06$\pm$0.02, consistent 
with the value measured toward compact \HII\ regions in the Galaxy. 

\item
No $He^{++}$ RRLs were detected toward the observed sources (1 $\sigma$
upper limit is 4 mK), which ruled out the possibility that helium
may be doubly ionized in the diffuse regions of the \UCHII\ envelopes. 

\item
Toward G10.15-0.34 \& G23.46-0.20, we investigated the 
spectrum of the radiation that ionizes the diffuse regions. 
We considered that the stars of type O6 and earlier
were embedded in (locally) ionization bounded regions and were not
contributing to the ionization of the diffuse regions.
The observed upper limit on the $n_{He^+}/n_{H^+}$ ratio then provided
an upper mass of $\sim$ 30 \Msun for the cluster members that ionize the diffuse
regions. This upper mass, however, is somewhat uncertain due 
to statistical uncertainty in the cluster population. 

\item
Our investigation ruled out accreting low-mass stars \citep{smith14,cesetal16} 
as a possible source for additional hydrogen ionizing photons required to produce
low $n_{He^+}/n_{H^+}$ ratio in the \UCHII\ envelopes.  

\item
We also investigated whether selective absorption of Lyman photons 
by dust was responsible for low helium ionization. We found that 
the ratio of helium and hydrogen absorption cross section of 
Lyman photons by dust, $a_0$, should be in the range 1.8 to 4.4 
to account for the observed $n_{He^+}/n_{H^+}$ ratio. However,
a self-consistent model for dust absorption cross section over
a wide wavelength range needs to be developed to conclusively
establish the role of selective absorption. 
\end{enumerate} 

\appendix

\section{Appendix: Computation of $\gamma$ and $a_0$}

Here we summarize the computation of (a) $\gamma$, the helium to hydrogen
photon number ratio, due to a star cluster; (b) $\gamma$ 
with emission from low-mass accreting stars in the cluster and (c)
the parameter $a_0$ which characterize the selective dust absorption of
Lyman photons.

\subsection{Expected helium ionization due to a cluster}
\label{a1}

Computation of hydrogen ($q_{H}$) and helium ($q_{He}$) Lyman photons from a cluster requires
a knowledge of these photon emission by stars of different mass and
the mass function of the cluster. We consider here a modified version of Muench IMF
\citep{muenchetal02} discussed by \citet{mr10} 
\bea
\zeta(m) \equiv  m\; dN/dm  & = & N_0\; m^{-\Gamma} \qquad\qquad\qquad\qquad\qquad\qquad\qquad  m_u > m > m_1 \\
         & = & N_0\; m_1^{(0.15-\Gamma)}\; m^{-0.15} \qquad\qquad\qquad\qquad\;\;\;\   m_1 > m > m_2 \\
         & = & N_0\; m_1^{(0.15-\Gamma)}\; m_2^{(-0.73-0.15)} \; m^{0.73} \qquad\qquad\; m_2 > m > m_l 
\eea
We used $\Gamma = 1.35$, the slope of Salpeter IMF, for high mass stars
\citep{mr10}, $m_l = 0.1$ \Msun, $m_1 = 0.6$ \Msun ($m_2 = 0.025$ \Msun,
which is below $m_l$). The maximum value for the upper mass cutoff, $m_u$,
is taken as 120 \Msun. $N_0$ is fixed through the normalization
\be
\int_{m_l}^{m_u} \zeta(m) dm/m = 1
\ee 
The Lyman photon emission from stars are taken from the models
of \citet{msh05}. With the above IMF and $m_u =120$ \Msun, 
we get a mean mass of $<m> = 0.71$ \Msun and mean Lyman continuum photons per
sec per \Msun, $<q>/<m> = 6.1 \times 10^{46}$ s$^{-1}$ \Msun$^{-1}$. These 
values are consistent with those obtained by \citet{mr10}. The helium to hydrogen
Lyman photon number ratio is
\be
\gamma \approx <q_{He}>/<q_{H}>
\label{gamma}
\ee
where $<q_{He}>$ and $<q_{H}>$ are the IMF weighted helium and hydrogen
Lyman photon emission per sec. $<q_{He}>$ is approximately taken
as the Lyman continuum photons with $\lambda < 504$ \AA~\citep{msh05}.
$\gamma$ is computed for different $m_u$ in the range 15 to 120 \Msun.
The result of the computation is plotted in Fig.~\ref{fig9}.

\subsection{The effect of accreting low mass stars}
\label{a2}

Several \HII\ regions have Lyman continuum emission in excess of
what is expected from the star type determined from their IR emission
\citep{wc89,sanetal13}. A model that can account for 
the excess Lyman photon emission in terms of accreting low-mass
stars is presented by \citet{smith14} (see also 
\nocite{hyo10}Hosokawa, Yorke \& Omukai 2010). Recent 
observation of infall tracers toward Lyman excess \HII\ regions
supports the accretion model \citep{cesetal16}.
Here we use Fig.~2 and Fig.~3 of \citet{smith14}, which gives 
respectively the stellar radius and accretion luminosity as 
a function of the mass of the accreting star. We consider
the `cold' accretion case as suggested by \citet{smith14}. 
The mass-radius and mass-luminosity relationships are 
given for different accretion rates. To account
for the observed Lyman excess in \HII\ regions
accretion rates in the range $10^{-5}$ to 10$^{^-3}$ \Msun yr$^{-1}$
are required. Calculations here are presented with accretion rate 
of 10$^{-4}$ \Msun yr$^{-1}$. The excess Lyman emission
originates at `hot' spots created on the surface of the star
during `cold' accretion. The temperature, $T_{h}$, 
of the hot spot is obtained by assuming a fraction, $f_{acc}$,
of the accretion luminosity is converted to thermal energy.
Thus 
\be
T_{h}^4 = \frac{f_{acc} L_{acc}}{4\pi R_s^2 f_{h} \sigma} 
\ee
where $L_{acc}$ is the accretion luminosity, $\sigma$ is
the Stefan-Boltzmann constant, $R_s$ is the radius of the central star,
$f_{h}$ is the fraction of the surface area covered by the
hot spot. Following \citet{smith14}, we take $f_{acc} = 0.75$ and
$f_{h} = 0.05$. The Lyman continuum photon emission from this 
hot spot is taken as that due to a star with
the same effective temperature in the models of \citet{msh05}.
The fraction of low-mass stars in cold accretion phase
with hot spot in the cluster is not known;
we assume 1 \% and 5\% of low-mass stars are in such a phase for 
the computation. We re-computed $\gamma$ of the radiation
from the cluster by considering 1 \% and 5\% of 
stars in the mass range 1 to 6 \Msun are in the accretion phase
and have hot spots.
The result of these computations is shown in Fig.~\ref{fig9}. 

\subsection{The effect of dust within the \HII\ region}
\label{a3}

We are interested in the effect of dust within the \HII\ region 
on the size of the helium ionization region. In Section~\ref{sec:dis}, we
presented a physical argument to illustrate that in a dust free
\HII\ region the size of the ionization zones of helium and
hydrogen is determined by the Lyman photon number ratio $\gamma$.
In ionization equilibrium equation, the zone sizes are involved
to get the total recombination rates of hydrogen and helium, which are volume integrals
over the respective ionization regions. The observed helium to hydrogen ratio 
are obtained from the ratio of
antenna temperatures (see Eq.~\ref{lineanttemp}). The antenna temperatures
depend on the line flux densities and hence are proportional to the total 
recombination rates. Thus
\be
\frac{n_{He^+}}{n_{H^+}} = \frac{\int_{V(He^+)} n_{He^+} n_e dV}{\int_{V(H^+)} n_{H^+} n_e dV} \approx y R.
\label{ratioandR}
\ee
where
\be
R \equiv \frac{\int_{V(He^+)} n_{H^+}^2 dV}{\int_{V(H^+)} n_{H^+}^2 dV} \approx \frac{R_{He^+}^3}{R_{H^+}^3},
\label{eqR}
\ee
the volume integrals in the numerator
and denominator are over the regions, respectively, where helium and hydrogen 
are ionized, $R_{He^+}$ and $R_{H^+}$ are radii of the two zones, 
if they are spherical in shape,
$n_{He^+}$ is the helium ion density, $n_{H^+}$ is the proton density, 
$n_e$ is the electron density and $y$ is the cosmic abundance of helium. 
\citet{mathis71} derived numerically the
relationship between $\gamma$ and $R$ for $y = 0.1$ (Fig. 3 in \citet{mathis71}; see also comment
by \citet{msc74}).

The dust in the \HII\ region absorbs and scatters Lyman photons.
Let $\sigma_H$ and $\sigma_{He}$ be the effective (i.e. weighted by stellar 
radiation flux density) absorption cross section of
hydrogen and helium Lyman photons respectively, assumed to be 
constant over the corresponding wavelength range. The absorption
optical depths of hydrogen and helium Lyman photons are  
\bea
\tau_H & = & n_{H^+} \sigma_H x_g R_{H^+} \phi^{1/2} \nonumber \\
       & \approx & 1.8 \times 10^{-21} \frac{\sigma_H x_g}{\sigma_{WD} x_g/(1-\Gamma)} n_e R_{H^+} 
                                                                       \phi^{1/2} \nonumber \\ 
       &  =  & 1.8 \times 10^{-21} f_{uv} n_e R_{H^+} \phi^{1/2} 
\label{dustodh}
\eea
\bea
\tau_{He} & = & n_{H^+} \sigma_{He} x_g R_{He^+} \phi^{1/2} \nonumber \\
       & \approx & 1.8 \times 10^{-21} a_0 f_{uv} n_e R_{H^+} \phi^{1/2}
\label{dustodhe}
\eea
where $x_g$ is the dust to gas number density ratio, $\phi$ is
the filling factor and the parameter $a_0 \equiv \sigma_{He}/\sigma_H$.  
The dust model of \citet{wd01}, derived for diffuse ISM, 
provides the extinction cross section, $\sigma_{WD}$, 
near 13.6 eV as $\sim$ 1.8 $\times 10^{-21}$ cm$^2$/H-atom
(their Fig. 14, $R_V$ = 3.1 model).
In Eq.~\ref{dustodh} \& \ref{dustodhe}, we expressed the absorption optical depths
in terms of the \citet{wd01} absorption cross section at 13.6 eV, which
is $\sigma_{WD}/(1-\Gamma)$ where $\Gamma$ is the dust albedo. 
We assume $\Gamma$ is constant over the Lyman continuum. 
A parameter $f_{uv} \equiv \frac{\sigma_H x_g}{\sigma_{WD} x_g/(1-\Gamma)}$  
is defined to take into account of deviations from the \citet{wd01} dust model.
The absorption due to dust changes the Lyman photon rates, which can 
be expressed as 
\bea
q_H^{'} & = & q_H \textrm{exp}(-\tau_H)  \\
q_{He}^{'} & = & q_{He} \textrm{exp}(-\tau_{He}). 
\eea  
If $\tau_{He}$ is not equal to $\tau_{H}$, the two photon rates 
change differently, which results in selective dust absorption proposed by \citet{msc74}
(see also \nocite{ps78}Panagia \& Smith 1978). Thus selective absorption
modifies Lyman photon ratio, $\gamma' = \frac{q_H^{'}}{q_{He}^{'}}$, which
in turn affects the sizes of the ionization zones of the two atoms \citep{mathis71}.  
$\gamma'$ is related to the original photon rate $\gamma$ as \citep{msc74, ps78} 
\be
\gamma' = \gamma \textrm{exp}(-(\tau_{He} - \tau_{H})) = \gamma \textrm{exp}(-\tau_H (a_0 R^{1/3} - 1)) 
\label{gammadash}
\ee
where the ratio $R_{He^+}/R_{H^+} = R^{1/3}$ for spherically symmetric
ionization zones but is a good approximation for non-uniform \HII\ regions
\citep{msc74, ps78}. Note that, if the ionization is due to a stellar cluster then
in the above equation $\gamma$ and hence $\gamma'$ needs to be obtained as given 
in Eq.~\ref{gamma} 

Our aim here is to get values for $a_0$ from the observed $n_{He^+}/n_{H^+}$
values or its upper limit. We use Eq.~\ref{ratioandR} to get $R$ from
the observed value $n_{He^+}/n_{H^+}$ for $y \approx 0.1$. 
If selective dust absorption is present, then 
Fig. 3 of \citet{mathis71} will provide a $\gamma'$ corresponding to the estimated
$R$. The $\gamma$ for the cluster radiation is 
taken as 0.2 (see Fig.~\ref{fig9}), a reasonable value
for the sources we have observed. Eq.~\ref{gammadash} \& \ref{dustodh} can then be used to get $a_0$
for a given $\phi$ and $f_{uv}$. The radius $R_{H^+}$
is estimated from the angular size and distance to the sources provided in Table~\ref{tab1}
and $n_e$ is obtained using the Lyman continuum luminosity given in Table~\ref{tab1} and $R_{H^+}$. 
The estimated $a_0$ values for a set of $\phi$ and $f_{uv}$ are given in Table~\ref{tab5}.
   
\acknowledgments
We thank Drs Kim and Koo for providing their 21cm images. We acknowledge the critical
comments by an anonymous referee that have significantly improved the presentation
of the results in the paper.

Facility: \facility{Green Bank Telescope}


\begin{deluxetable}{lrr} 
\tabletypesize{\footnotesize}
\tablecolumns{3}
\tablewidth{0pc}
\tablecaption{Properties of the selected sources \label{tab1}}
\tablehead{
\colhead{Properties} & \colhead{Value} & \colhead{Ref/Note}  
 } 
\startdata 
\cutinhead{G10.15$-$0.34}
Distance & 3.6 kpc & 1 \\
\UCHII\ regions & G10.15$-$0.34 & 2 \\
Angular size of envelope & 10\arcmin.9 $\times$ 6\arcmin.7 & 3 \\
Linear size of envelope\tablenotemark{a} & 11.3 pc $\times$ 6.9 pc &  \\
Flux density of envelope at 1.43 GHz & 55.22 Jy & 3 \\
Assumed electron temperature of the ionized gas & 8000 K &  \\
Lyman continuum luminosity\tablenotemark{b}    & 6.0 $\times 10^{49}$ s$^{-1}$ &  \\ 
Star type\tablenotemark{c} & O3 V & \\ 
\cutinhead{G23.46$-$0.20}
Distance & 6 kpc & 4,5 \\
\UCHII\ regions & G23.455$-$0.201& 2 \\
Angular size of envelope & 8\arcmin.8 $\times$ 5\arcmin.8 & 3 \\
Linear size of envelope\tablenotemark{a} & 15.4 pc $\times$ 10.1 pc &  \\
Flux density of envelope at 1.43 GHz & 11.31 Jy & 3 \\
Assumed electron temperature of the ionized gas & 8000 K &  \\
Lyman continuum luminosity\tablenotemark{b} & 3.5 $\times 10^{49}$ s$^{-1}$ & \\ 
Star type\tablenotemark{c} & O4.5 V &  \\ 
\cutinhead{G29.96$-$0.02}
Distance & 6.2 kpc & 6  \\
\UCHII\ regions & G29.95$-$0.01, G29.86$-$0.04 & 2, 6 \\
Angular size of envelope & 6\arcmin.3 $\times$ 5\arcmin.2 & 3 \\
Linear size of envelope\tablenotemark{a} & 11.7 pc $\times$ 9.4 pc &  \\
Flux density of envelope at 1.43 GHz & 12.69 Jy & 3 \\
Assumed electron temperature of the ionized gas & 8000 K &  \\
Lyman continuum luminosity\tablenotemark{b} & 4.2 $\times 10^{49}$ s$^{-1}$ &  \\ 
Star type\tablenotemark{c} & O4 V & \\ 
\enddata
\tablerefs{
(1)\citet{uetal12}, (2)\citet{wc89}, (3)\citet{kimkoo01}, 
(4) \citet{wienenetal15}, (5) \citet{brunetal09},
(6)\citet{zhangetal14}} 
\tablenotetext{a}{ Linear size is estimated using the distance give in the table here}
\tablenotetext{b}{ Lyman continuum luminosity is estimated as described in \citet{rubin68} 
using the flux density at 
1.43 GHz, distance to the source and electron temperature given in the table here.
No correction for dust extinction is applied.}
\tablenotetext{c}{Type of the star from the estimated Lyman continuum luminosity using
the results of \citet{msh05}.} 
\end{deluxetable} 

\begin{deluxetable}{lrrrlr} 
\tabletypesize{\small}
\tablecolumns{6}
\tablewidth{0pc}
\tablecaption{Summary of Observation \label{tab2}}
\tablehead{
\colhead{Source} & \colhead{RA(2000)} & \colhead{DEC(2000)} & \colhead{Obs. Time} & \colhead{RRLs Averaged} & \colhead{Eff. Int.} \\ 
\colhead{} &  &  & \colhead{(minutes)} &  & \colhead{(hrs)} 
 } 
\startdata 
\cutinhead{G10.15$-$0.34} 
G10.15$-$0.34a &  18:09:23.5 & $-$20:19:25 &  4.9 & 104, 109, 110, 111, 113 &  0.8 \\ 
G10.15$-$0.34b &  18:09:09.5 & $-$20:21:00 & 14.6 & 104, 109, 110, 111 &  2.0 \\ 
G10.15$-$0.34c &  18:09:23.5 & $-$20:22:25 &  9.8 & 104, 109, 110, 111 &  1.3 \\ 
G10.15$-$0.34d &  18:09:36.0 & $-$20:22:25 &  9.8 & 104, 109, 110, 111 &  1.3 \\ 
\cutinhead{G23.46$-$0.20} 
G23.46$-$0.20a &  18:34:44.7 & $-$08:32:17 &  7.3 & 104, 109, 110, 111, 113 &  1.2 \\ 
G23.46$-$0.20b &  18:34:32.7 & $-$08:32:17 &  9.8 & 104, 109, 110, 111 &  1.3 \\ 
G23.46$-$0.20c &  18:34:24.7 & $-$08:33:60 &  9.8 & 104, 109, 110, 111 &  1.3 \\ 
G23.46$-$0.20d &  18:35:03.7 & $-$08:33:09 &  9.8 & 104, 109, 110, 111 &  1.3 \\ 
G23.46$-$0.20e &  18:35:13.7 & $-$08:33:08 &  9.8 & 104, 109, 110, 111 &  1.3 \\ 
G23.46$-$0.20f &  18:35:40.6 & $-$08:30:21 &  9.8 & 104, 109, 110, 111 &  1.3 \\ 
\cutinhead{G29.96$-$0.02} 
G29.96$-$0.02a &  18:46:10.4 & $-$02:41:45 &  4.9 & 104, 109, 110, 111, 113 &  0.8 \\ 
G29.96$-$0.02b &  18:45:56.8 & $-$02:42:16 &  9.8 & 104, 109, 110, 111 &  1.3 \\ 
G29.96$-$0.02c &  18:46:15.8 & $-$02:38:15 &  9.8 & 104, 109, 110, 111 &  1.3 \\ 
G29.96$-$0.02d &  18:46:20.8 & $-$02:40:14 &  9.8 & 104, 109, 110, 111 &  1.3 \\ 
G29.96$-$0.02e &  18:46:21.8 & $-$02:37:44 &  9.8 & 104, 109, 110, 111 &  1.3 \\ 
G29.96$-$0.02f &  18:46:04.9 & $-$02:45:45 &  8.1 & 104, 109, 110, 111 &  1.1 \\ 
\enddata 
\end{deluxetable} 

\begin{deluxetable}{lrrrl} 
\tabletypesize{\small}
\tablecolumns{5}
\tablewidth{0pc}
\tablecaption{Observed line parameters \label{tab3}}
\tablehead{
\colhead{Source} & \colhead{T$_L$}\tablenotemark{a} & \colhead{$\Delta V$}\tablenotemark{a} 
 & \colhead{$V_{LSR}$}\tablenotemark{a} & \colhead{Line}  \\ 
\colhead{} & \colhead{(K)} & \colhead{(km s$^{-1}$)} & 
 \colhead{(km s$^{-1}$)} &   
 } 
\startdata 
G10.15-0.34a & 2.071 (0.034) & 27.38 (0.2) & 13.03 (0.03) & H \\ 
  & 0.972 (0.035) & 51.68 (0.6) & 13.17 (0.09) & H  \\ 
  & 0.116 (0.018) & 27.42 (1.1) & 13.03 (0.00) & He  \\ 
  & 0.056 (0.005) & 34.46 (6.3) & 21.33 (3.86) & C\tablenotemark{b}  \\ 
G10.15-0.34b & 0.047 (0.001) & 36.98 (0.4) & 16.13 (0.18) & H \\ 
   & $\le$ 0.002 &  &  & He, C  \\
G10.15-0.34c & 0.048 (0.004) & 24.47 (1.3) & 12.77 (0.34) & H \\ 
  & 0.071 (0.005) & 48.06 (1.0) &  6.29 (0.46) & H  \\ 
   & $\le$ 0.003 &  &  & He, C  \\
G10.15-0.34d & 0.067 (0.003) & 19.15 (0.6) &  8.46 (0.18) & H \\ 
  & 0.092 (0.003) & 43.56 (0.5) & 16.34 (0.30) & H  \\ 
  & 0.006 (0.001) & 14.86 (2.7) &  8.46 (0.00) & He  \\ 
   & $\le$ 0.003 &  &  & C  \\
G10.15-0.34avg\tablenotemark{c} & 0.064 (0.001) & 38.98 (0.3) & 12.83 (0.13) &  H \\
   & $\le$ 0.002 &  &  & He,C \\
G23.46-0.20a & 0.547 (0.001) & 24.08 (0.1) & 101.06 (0.04) & H \\ 
  & 0.040 (0.001) & 30.26 (1.4) & 64.60 (0.58) & H  \\ 
  & 0.033 (0.001) & 19.41 (1.1) & 101.06 (0.00) & He  \\ 
  & 0.015 (0.002) &  9.50 (1.6) & 102.50 (0.68) & C  \\ 
G23.46-0.20b & 0.289 (0.001) & 22.75 (0.1) & 99.10 (0.06) & H \\ 
  & 0.025 (0.001) & 22.95 (1.6) & 57.32 (0.65) & H  \\ 
  & 0.013 (0.001) & 15.38 (2.0) & 99.10 (0.00) & He  \\ 
  & 0.009 (0.002) &  7.34 (2.0) & 102.10 (0.87) & C  \\ 
G23.46-0.20c & 0.233 (0.001) & 19.48 (0.1) & 98.10 (0.05) & H \\ 
  & 0.022 (0.001) & 27.87 (1.7) & 58.17 (0.67) & H  \\ 
  & 0.013 (0.001) & 12.44 (1.4) & 98.10 (0.00) & He  \\ 
   & $\le$ 0.003 &  &  & C  \\
G23.46-0.20d & 0.040 (0.001) & 15.30 (0.7) & 57.03 (0.36) & H \\ 
  & 0.026 (0.001) & 23.97 (1.7) & 99.94 (0.86) & H  \\ 
  & 0.016 (0.001) & 17.54 (3.3) & 76.78 (0.97) & H  \\ 
   & $\le$ 0.002 &  &  & He, C  \\
G23.46-0.20e & 0.045 (0.001) & 16.87 (0.7) & 59.15 (0.32) & H \\ 
  & 0.035 (0.001) & 26.24 (1.7) & 95.61 (1.00) & H  \\ 
  & 0.021 (0.003) & 14.17 (2.0) & 77.70 (0.66) & H  \\ 
   & $\le$ 0.003 &  &  & He, C  \\
G23.46-0.20f & 0.049 (0.001) & 26.47 (0.5) & 89.20 (0.19) & H \\ 
  & 0.014 (0.001) & 17.82 (1.3) & 57.34 (0.55) & H  \\ 
   & $\le$ 0.002 &  &  & He, C  \\
G23.46-0.20avg\tablenotemark{d}& 0.026 (0.001) & 15.89 (0.5)&  56.87 (0.21) &  H \\
   &0.035 (0.001) &  32.32 (0.6) &  90.36 (0.22) &  H \\
   & $\le$ 0.002 &  &  & He, C \\
G29.96-0.02a & 0.512 (0.016) & 29.11 (0.4) & 101.35 (0.11) & H \\ 
  & 0.083 (0.014) & 46.37 (1.5) & 89.35 (2.21) & H  \\ 
  & 0.045 (0.002) & 26.94 (1.1) & 101.35 (0.00) & He  \\ 
  & 0.018 (0.004) &  4.97 (1.1) & 99.67 (0.48) & C  \\ 
G29.96-0.02b & 0.461 (0.005) & 19.85 (0.1) & 99.28 (0.04) & H \\ 
  & 0.054 (0.005) & 40.26 (1.3) & 93.58 (0.62) & H  \\ 
  & 0.033 (0.001) & 19.18 (0.7) & 99.28 (0.00) & He  \\ 
  & 0.014 (0.001) & 10.11 (1.2) & 97.09 (0.51) & C  \\ 
G29.96-0.02c & 0.180 (0.005) & 18.23 (0.3) & 95.45 (0.07) & H \\ 
  & 0.035 (0.005) & 43.33 (2.6) & 95.25 (0.56) & H  \\ 
  & 0.013 (0.001) & 13.55 (1.6) & 95.45 (0.00) & He  \\ 
   & $\le$ 0.003 &  &  & C  \\
G29.96-0.02d & 0.141 (0.003) & 19.11 (0.3) & 96.02 (0.07) & H \\ 
  & 0.040 (0.003) & 46.11 (1.7) & 95.91 (0.38) & H  \\ 
  & 0.010 (0.001) & 15.94 (1.6) & 96.02 (0.00) & He  \\ 
   & $\le$ 0.003 &  &  & C  \\
G29.96-0.02e & 0.166 (0.002) & 20.07 (0.2) & 96.89 (0.07) & H \\ 
  & 0.014 (0.002) & 56.08 (3.7) & 87.29 (1.85) & H  \\ 
  & 0.012 (0.001) & 15.57 (1.5) & 96.89 (0.00) & He  \\ 
   & $\le$ 0.003 &  &  & C  \\
G29.96-0.02f & 0.283 (0.002) & 19.67 (0.2) & 97.68 (0.07) & H \\ 
  & 0.023 (0.002) & 14.50 (1.4) & 97.68 (0.00) & He  \\ 
  & 0.007 (0.003) &  8.25 (3.5) & 102.62 (1.47) & C  \\ 
\enddata 
\tablenotetext{a}{1$\sigma$ errors in the estimated quantities
are given in the bracket. The upper limit on line temperature
are 1$\sigma$ limit. If a parameter is fixed during Gaussian fitting,
then the corresponding error is set to 0.0.}
\tablenotetext{b}{This line component is tentatively identified as due to carbon atom based
on its LSR velocity offset from the hydrogen RRL. However, we note the larger line width compared to the
typical width (\lsim 10 \kms) of carbon lines observed toward PDRs.}
\tablenotetext{c}{Spectrum obtained by averaging the data toward positions
G10.15-0.34b and G10.15-0.34c.}
\tablenotetext{d}{Spectrum obtained by averaging the data toward positions
G23.46-0.20d and G23.46-0.20f.}
\end{deluxetable} 


\begin{deluxetable}{lrrrrl} 
\tabletypesize{\small}
\tablecolumns{6}
\tablewidth{0pc}
\tablecaption{$n_{He^+}/n_{H^+}$ ratio toward the observed positions \label{tab4}}
\tablehead{
\colhead{Source} & \colhead{$n_{He^+}/n_{H^+}$}\tablenotemark{a} & \colhead{$V_{LSR}$} & \colhead{Offset} & 
\colhead{T$_{b4.9}$}\tablenotemark{b} & \colhead{EM}  \\ 
\colhead{} & \colhead{} & \colhead{(km s$^{-1}$)} & \colhead{(pc)} & \colhead{(K)} &  \colhead{(pc cm$^{-6}$)}          
 } 
\startdata 
G10.15-0.34a  &  0.056 (0.009)  & 13.03      &  0.9 & 50.0 &  $4.0 \times 10^5$\\
G10.15-0.34b  &  $\le$   0.052  & 16.13      &  4.7 &  5.0 &  $4.0 \times 10^4$\\
G10.15-0.34c  &  $\le$   0.037  & 12.77,6.29 &  3.1 &  5.0 &  $4.0 \times 10^4$\\
G10.15-0.34d  &  0.073 (0.018)  & 8.46       &  3.9 &  5.0 &  $4.0 \times 10^4$\\
G10.15-0.34avg&  $\le$   0.033  & 12.83      &  3.9 &  \nodata    &  $4.0 \times 10^4$\\
G23.46-0.20a  &  0.048 (0.003)  & 101.06     &  2.0 &  8.0 &  $6.4 \times 10^4$\\
G23.46-0.20b  &  0.031 (0.005)  & 99.10      &  5.7 &  6.0 &  $4.8 \times 10^4$\\
G23.46-0.20c  &  0.035 (0.005)  & 98.10      & 10.1 &  3.6 &  $2.9 \times 10^4$\\
G23.46-0.20d  &  $\le$   0.094  & 99.94      &  9.0 &  2.0 &  $1.6 \times 10^4$\\
G23.46-0.20e  &  $\le$   0.072  & 95.61      & 13.0 &  1.8 &  $1.4 \times 10^4$\\
G23.46-0.20f  &  $\le$   0.050  & 89.20      & 24.4 &  1.1 &  $8.7 \times 10^3$\\
G23.46-0.20avg&  $\le$   0.051  & 90.36      & 15.5 &  \nodata & $1.2 \times 10^4$\\
G29.96-0.02a  &  0.081 (0.005)  & 101.35     &  5.2 &  10.0 & $7.9 \times 10^4$\\
G29.96-0.02b  &  0.069 (0.003)  & 99.28      &  6.1 &   8.0 & $6.4 \times 10^4$\\
G29.96-0.02c  &  0.053 (0.008)  & 95.45      &  5.7 &   4.0 & $3.2 \times 10^4$\\
G29.96-0.02d  &  0.061 (0.008)  & 96.02      &  7.8 &   2.8 & $2.2 \times 10^4$\\
G29.96-0.02e  &  0.058 (0.007)  & 96.90      &  8.6 &   2.4 & $1.9 \times 10^4$\\
G29.96-0.02f  &  0.061 (0.008)  & 97.68      & 11.5 &   3.0 & $2.4 \times 10^4$\\
\enddata 
\tablenotetext{a}{1$\sigma$ error in the estimated quantity
is given in the bracket. The upper limits are also 1$\sigma$ limit}
\tablenotetext{b}{Main beam brightness temperature at 4.875 GHz at the observed positions
obtained from the galactic plane survey of \citet{aetal79}. 
}
\end{deluxetable} 

\begin{deluxetable}{lrr} 
\tabletypesize{\footnotesize}
\tablecolumns{3}
\tablewidth{0pc}
\tablecaption{Selective absorption due to dust \label{tab5}}
\tablehead{
\colhead{Properties} & \colhead{Value} & \colhead{Note}  
 } 
\startdata 
\cutinhead{G10.15$-$0.34}
Path length (pc)               &   4.4                         & \tablenotemark{a}           \\ 
Electron density (\cmthree)    &   135                         & \tablenotemark{b}           \\
$n_{He^+}/n_{H^+}$             &   $\le$ 0.033                  & From Table~\ref{tab4}       \\
$\gamma'$                      &   0.04\tablenotemark{c}                        &           \\
a0                             &   4.4  & $\phi$=0.25\tablenotemark{d}, $f_{uv}$=0.5\tablenotemark{d}           \\
                               &   2.4  & $\phi$=0.64, $f_{uv}$=1.0          \\
                               &   2.9  & $\phi$=0.25, $f_{uv}$=1.0          \\
                               &   2.2  & $\phi$=0.25, $f_{uv}$=2.0          \\
\cutinhead{G23.46$-$0.20}
Path length (pc)               &  6.2                          & \tablenotemark{a}           \\ 
Electron density (\cmthree)    &  62                           & \tablenotemark{b}           \\
$n_{He^+}/n_{H^+}$             &   $\le$ 0.051                  & From Table~\ref{tab4}       \\
$\gamma'$                      &  0.06\tablenotemark{c}         &             \\
a0                             &  4.0   & $\phi$=0.25, $f_{uv}$=0.5            \\
                               &  2.1   & $\phi$=0.64, $f_{uv}$=1.0            \\
                               &  2.7   & $\phi$=0.25, $f_{uv}$=1.0            \\
                               &  2.0   & $\phi$=0.25, $f_{uv}$=2.0            \\
\cutinhead{G29.96$-$0.02}
Path length (pc)               &  5.2                          & \tablenotemark{a}           \\ 
Electron density (\cmthree)    &  90                           & \tablenotemark{b}           \\
$n_{He^+}/n_{H^+}$             &  0.053\tablenotemark{e}      & From Table~\ref{tab4}       \\
$\gamma'$                      &  0.06\tablenotemark{c}        &              \\
a0                             &  3.4   & $\phi$=0.25, $f_{uv}$=0.5             \\
                               &  1.9   & $\phi$=0.64, $f_{uv}$=1.0             \\
                               &  2.3   & $\phi$=0.25, $f_{uv}$=1.0             \\
                               &  1.8   & $\phi$=0.25, $f_{uv}$=2.0             \\
\enddata
\tablenotetext{a}{Path length, $R_{H^+}$, in Eq.~\ref{dustodh} estimated from the
angular size and distance given in Table~\ref{tab1} and assuming a spherical
distribution for the ionized gas \citep{pw78}.}
\tablenotetext{b}{Electron density, $n_e$, in Eq.~\ref{dustodh} estimated using the Lyman continuum 
luminosities given in Table~\ref{tab1} and $R_{H^+}$.}
\tablenotetext{c}{$\gamma'$ is obtained from the listed $n_{He^+}/n_{H^+}$ ratio as discussed in
Appendix~\ref{a3}.} 
\tablenotetext{d}{$\phi$ is the \HII\ region filling factor. $f_{uv}$ is the ratio of
the dust absorption cross section for hydrogen Lyman photons to the extinction cross section
at 13.6 eV in the \citet{wd01} dust model (see Appendix~\ref{a3}).} 
\tablenotetext{e}{Minimum $n_{He^+}/n_{H^+}$ ratio observed toward G29.96$-$0.02.}
\end{deluxetable}

\begin{figure}
\includegraphics[height=4.5in, width=7.0in, angle=0]{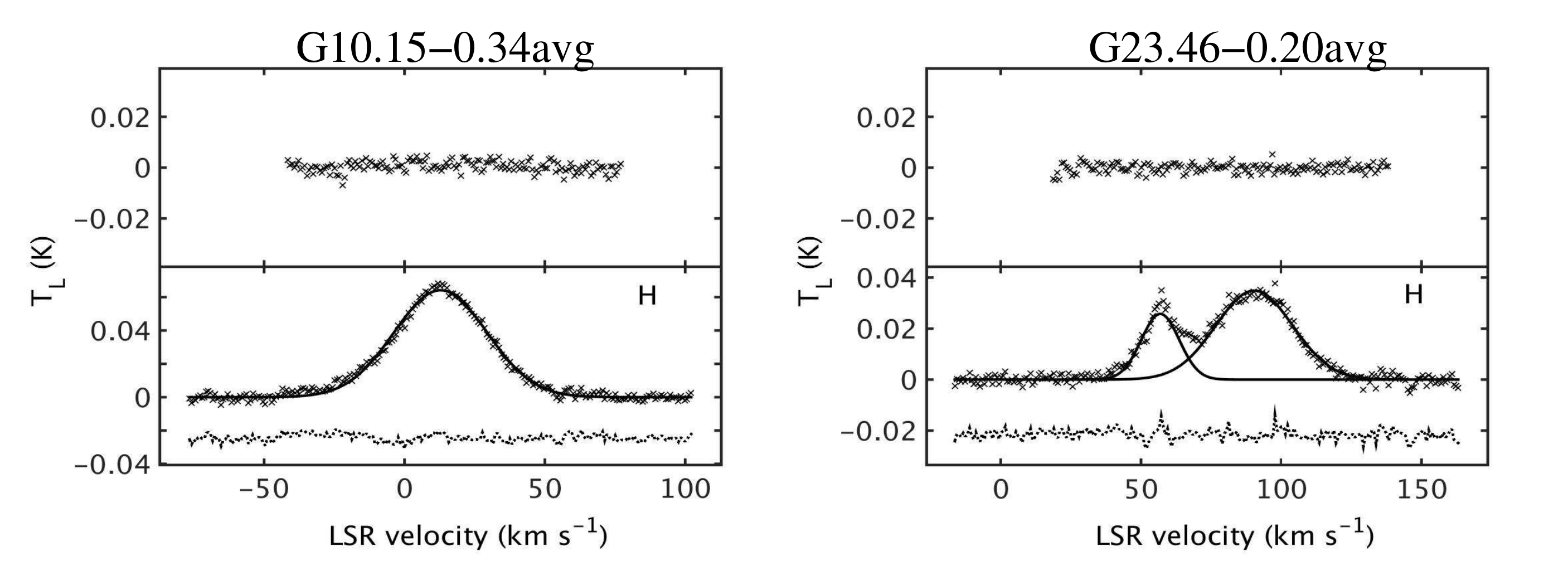}
\caption{
Spectra obtained by averaging the
data from positions G10.15-0.34b and G10.15-0.34c (left)
and from positions G23.46-0.20d and G23.46-0.20df (right).
The observed spectra (indicated in 'x'), the Gaussian
line components (continuous line) and the residual spectra (dotted line)
for hydrogen line are shown in the bottom panel. The helium spectra
are shown on the top panel.
\label{fig5} }
\end{figure}

\begin{figure}
\includegraphics[height=7in, angle=0]{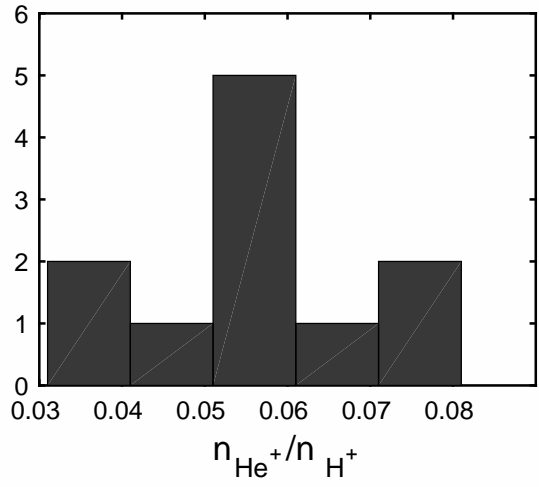}
\caption{
Histogram of the observed $n_{He^+}/n_{H^+}$ ratio.
\label{fig6} }
\end{figure}

\begin{figure}
\includegraphics[height=5.5in, width=5.0in, angle=0]{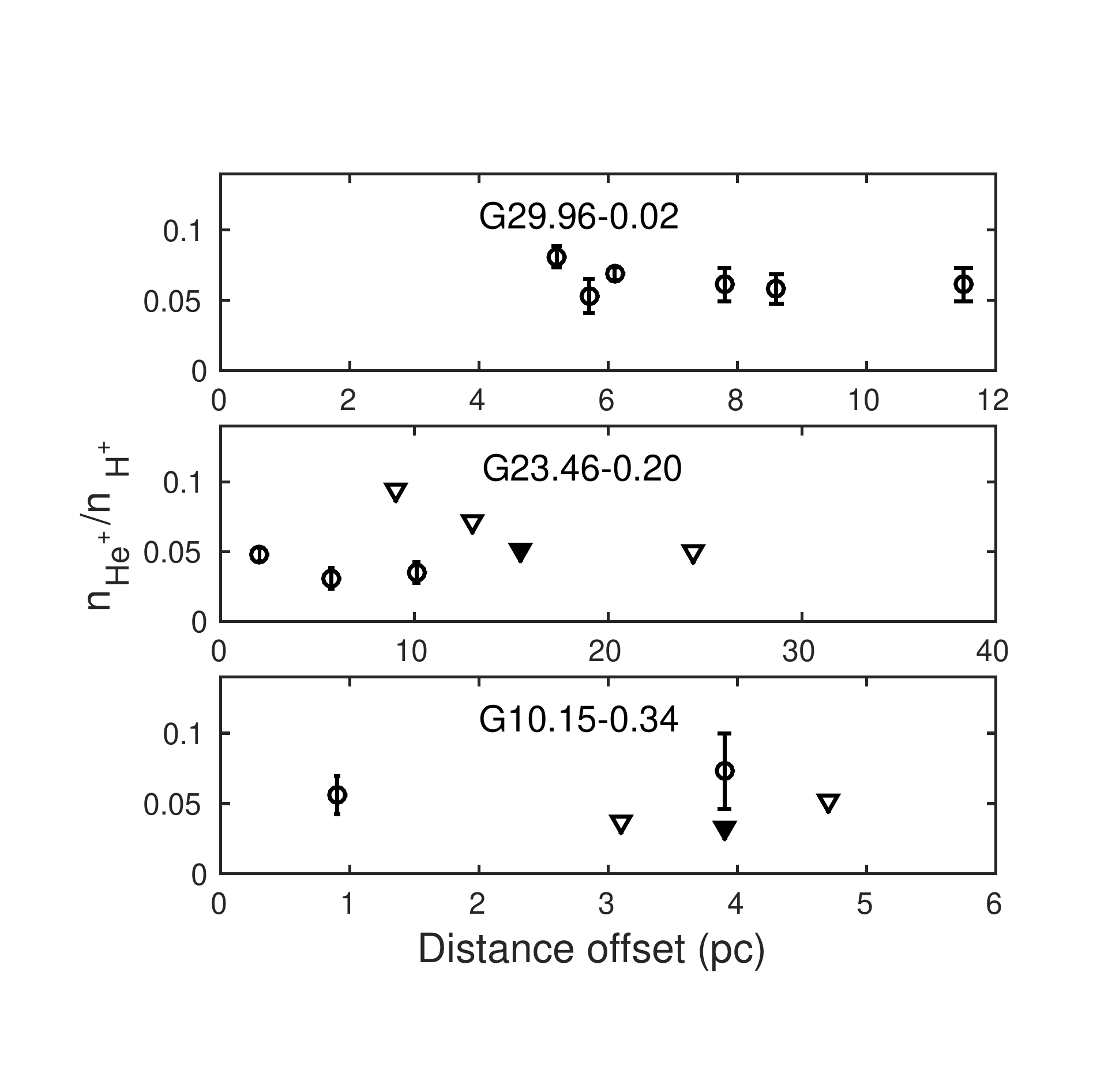}
\caption{
Observed $n_{He^+}/n_{H^+}$ ratio with $\pm$ 1.5 $\sigma$ error bar
vs distance from the `ionization
center' (see text). The $1\sigma$ upper limits on the ratio from non-detections 
are indicated by open triangles.  The filled triangles are $1\sigma$ upper limits 
obtained from the average spectra.
\label{fig7} }
\end{figure}

\begin{figure}
\includegraphics[height=5.5in, width=5.0in, angle=0]{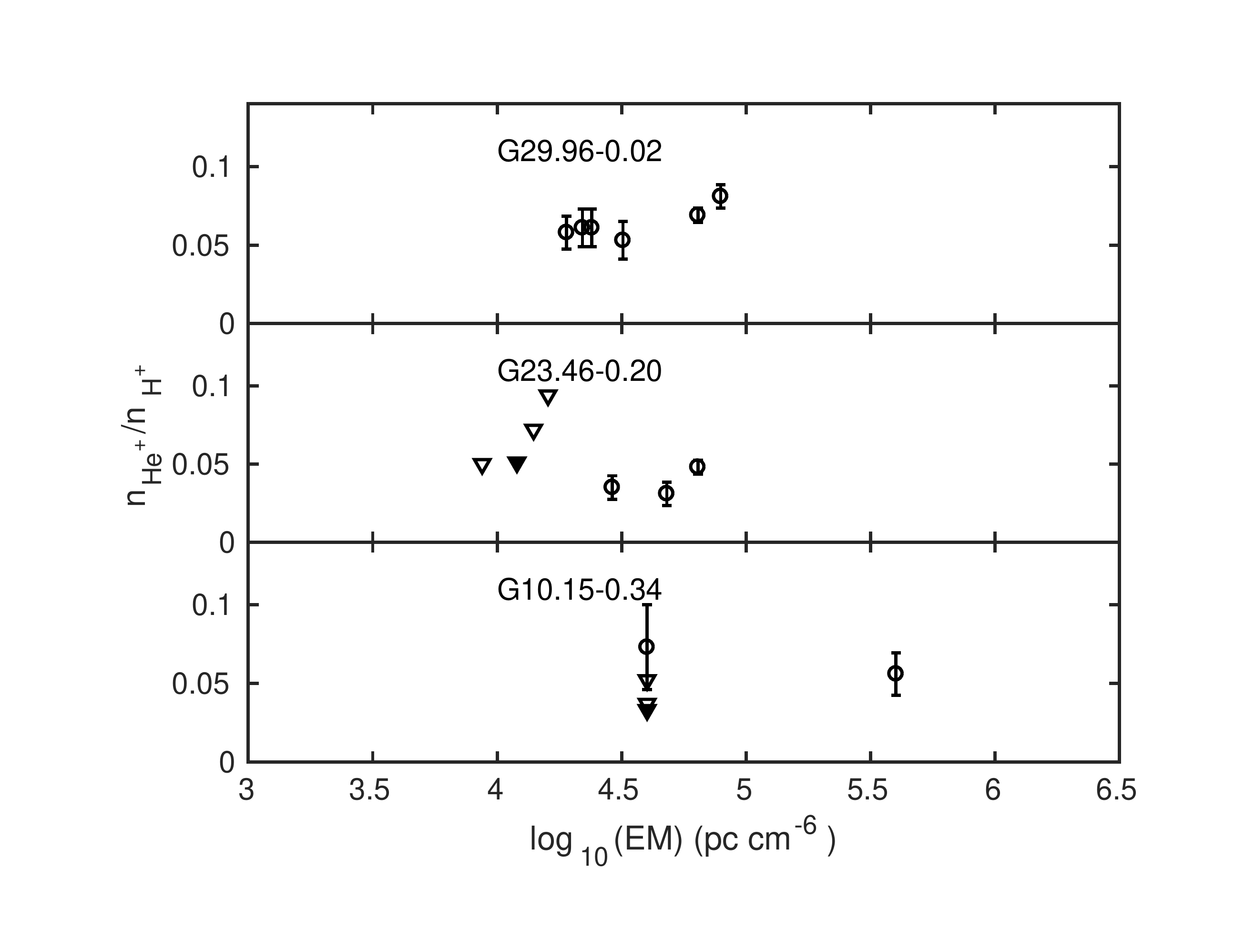}
\caption{
Observed $n_{He^+}/n_{H^+}$ ratio with $\pm$ 1.5 $\sigma$ error bar vs 
emission measure. The $1\sigma$ upper limits on the ratio from non-detections 
are indicated by open triangles.  The filled triangles are $1\sigma$ upper limits 
obtained from the average spectra. 
\label{fig8} }
\end{figure}

\begin{figure}
\includegraphics[height=5.0in, angle=0]{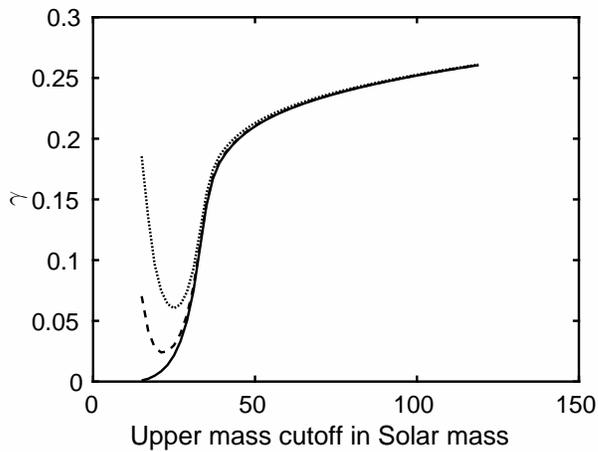}
\caption{
Helium to hydrogen Lyman photon ratio, $\gamma$,
vs upper cutoff mass of the cluster. The continuous curve shows $\gamma$
for ionizing radiation due to OB stars alone. A modified version
of Muench IMF is considered for the computation (see Appendix~\ref{a1}). 
Some of the cluster
members can be accreting, low-mass stars with `hot' spots, which are
suggested to be the possible source for excess Lyman emission 
observed in many galactic \HII\ regions \citep{smith14, cesetal16}. The
dashed and dotted curves are, respectively, the $\gamma$ computed by considering
that 1\% and 5\% of the stars in the mass range 1 to 6 \Msun are in the
accretion phase and contributing to the ionizing photon production rate
(see Appendix~\ref{a2}).  
\label{fig9} }
\end{figure}

\end{document}